# Reinforcement Learning Optimized Flight Control for Interception of High-Altitude Maneuvering Targets

Brian Gaudet*
*University of Arizona, 1127 E. Roger Way, Tucson Arizona, 85721*

**We implement and compare two approaches to improving flight control system (FCS) performance using reinforcement meta-learning (meta-RL). Both approaches are evaluated in a terminal phase hypersonic intercept engagement requiring hit-to-kill accuracy against a highly agile maneuvering target. In the first approach (FCS-RL), meta-RL is used to optimize a full FCS that maps acceleration tracking error and other measurements directly to commanded channel deflection rates of change, combining pitch, yaw, and roll control into a single policy. In contrast, the second approach (FCS-3L-RL) starts with a standard FCS (FCS-3L) implementing a three-loop autopilot for pitch and yaw control, but adapts the scheduled autopilot gains in real time using a policy optimized with meta-RL. Both approaches are compared to the FCS-3L in a 6-DOF simulator that captures most factors relevant to guidance and flight control. Simulations show that under ideal conditions the three systems have comparable performance. However, under conditions of aerodynamic uncertainty, radome refraction, or target evasive maneuvers, the FCS-RL significantly outperforms the FCS-3L-RL with improved intercept probability and less control effort. In turn, the FCS-3L-RL outperforms the FCS-3L by a similar margin. Importantly, we demonstrate that both FCS-RL and FCS-3L-RL generalize well to novel engagement scenarios, flight conditions, and target maneuvers that are not experienced during the optimization process.**

## I. Introduction

Flight control for control surface actuated interceptors tasked with intercepting actively evading targets in the high altitude hypersonic flight regime with hit-to-kill accuracy is a significant challenge. Consider that strike weapons fielded by a peer adversary will likely actively evade interception, requiring the interceptor to be capable of significant terminal phase maneuvering. These maneuvers can result in unsteady effects, boundary layer separation, and shock–boundary layer interaction [1, 2]. Aerodynamic modeling is further complicated by the extreme aero-thermal environment associated with hypersonic flight, where vehicle structures undergo stress caused by coupling of Fluid-Thermal-Structural Interactions (FTSI) and significant aerodynamic heating typical in hypersonic regimes [3]. This results in time-varying airframe and control surface deformation that are functions of both peak and integrated heating, which leads to vehicle aerodynamics differing substantially from that assumed during the design of the FCS [4, 5]. This aerodynamic uncertainty will require the flight control system (FCS) to adapt to differences between the simulated and deployment environments. The design of the FCS is further complicated by parasitic interaction between the guidance and flight control systems due to radome refraction [6–8]. Finally, high altitude engagements using only control surfaces actuators result in limited acceleration capability due to the low dynamic pressure.

Most currently deployed flight control systems for cruciform geometry skid-to-turn interceptors use the three-loop autopilot (TLA) [9–11] for pitch and yaw control, and a two-loop proportional-integral (roll rate and roll) controller for roll control. The TLA uses acceleration and rotational velocity measurements to track a commanded acceleration in the given channel (pitch and yaw). The FCS gains are optimized over a group of design points, typically angle of attack, sideslip angle, Mach, and altitude, and during flight the gains are interpolated between these design points to give optimal performance over the flight envelope. This approach gives both fast response and robust tracking to commanded acceleration. However, the standard TLA has limitations, including the assumptions of linear dynamics and no cross channel coupling, and simplified second order actuator model that does not account for fin limits and slew rate. Moreover, the three-loop architecture is a linear controller, which might constrain performance. Finally, as we show in the following, performance can degrade under high levels of aerodynamic uncertainty.

Consequently, considerable research has been focused on augmenting the standard three-loop autopilot, as well as exploring alternative approaches to interceptor flight control. Model-reference adaptive control (MRAC) methods

*Research Engineer, Department of Systems and Industrial Engineering, E-mail: briangaudet@arizona.edu

[12, 13] have been used to augment conventional three-loop autopilots through supplemental control commands that drive the missile acceleration response toward that of a desired reference model, improving tracking performance under aerodynamic model uncertainty. Despite these benefits, MRAC has several limitations. Conventional MRAC stability and tracking guarantees rely on plant-model matching conditions, which may be violated under substantial aerodynamic uncertainty or changes in control effectiveness [14]. Furthermore, the reference model may demand a response that exceeds the actuator constraints, resulting in saturation and degraded tracking performance [15]. Adaptive controllers can also be sensitive to unmodeled high-frequency dynamics and measurement disturbances, potentially compromising stability, particularly at high adaptation gains [16]. Finally, asymptotic tracking guarantees do not necessarily ensure satisfactory transient performance, motivating modified adaptive architectures that explicitly address transient response [17].

More recently, reinforcement learning (RL) has been investigated both for optimizing conventional missile autopilots and for replacing their prescribed feedback architectures. Shin et al. [18] employed deep reinforcement learning to optimize the gains of a conventional three-loop autopilot, preserving its underlying control structure while allowing state-dependent gain adaptation. Taligatos et al. [19] compared RL-based optimization of conventional controller gains against a fully learned controller that directly commands control surface deflections, highlighting tradeoffs in control effort versus overshoot between the two methods. Similarly, Candeli et al. [20] employed RL to directly synthesize a missile autopilot, demonstrating its effectiveness under nonlinear aerodynamic dynamics and model uncertainty. Wang et al. [21] extended this approach to integrated guidance and control for interception of highly maneuverable targets, with an RL policy directly commanding pitch and yaw control surfaces. Gaudet et al. [22] used a hypersonic point defense intercept application to compare the effectiveness of using reinforcement meta-learning (meta-RL) to optimize a separate FCS policy as compared to optimizing an integrated guidance and control policy, and in [23] investigated meta-RL for an integrated guidance and control system in an air-to-air intercept application. Collectively, these studies illustrate the potential of RL to improve conventional FCS performance.

In the following, we explore two approaches to apply reinforcement meta-learning (meta-RL) to optimize an adaptive FCS capable of hit-to-kill (miss less than 0.25 m) terminal-phase performance against maneuvering targets over a high-altitude hypersonic flight envelope using only control surface actuation. In the first approach (FCS-RL), meta-RL is used to optimize a full FCS mapping acceleration tracking error and other measurements directly to channel deflection rates of change, combining pitch, yaw, and roll control into a single policy. In contrast, the second approach (FCS-3L-RL) starts with a standard FCS (FCS-3L) implementing a three-loop autopilot for pitch and yaw control, but adapts the scheduled autopilot gains in real time using a policy optimized with meta-RL. This work has some similarity to [19] in that we explore using RL to augment a standard FCS as well as optimizing a full meta-RL FCS. However, our work uses a much higher fidelity simulator, uses proximal policy optimization in place of deep deterministic policy gradient, includes testing over full trajectories, evaluates performance under aerodynamic uncertainty and a large flight envelope, and uses meta-RL instead of RL. The recurrent policy used in our implementation of meta-RL allows the FCS-RL to implement an integrator, which we later show is essential for the RL optimized FCS to operate over a large flight envelope.

The paper is organized as follows. Section II describes the simulation environment and engagement scenarios, Section III covers the development of the FCS-3L benchmark, IV.A gives background for RL and meta-RL and Sections V and V.A formulate the FCS optimization problems in the meta-RL framework. Section VI describes the benchmark FCS and compares the performance of the meta-RL and benchmark flight control systems. Section VII gives a brief discussion of certification issues. Conclusions are given in Section VIII.

## II. Simulation Environment

In this section, unless a variable has the "B" (body frame) or "S" (LOS stabilized frame) superscript, it is assumed to be in the inertial reference frame. The simulator, taken from [24], has sufficient fidelity to capture most factors relevant to flight control. Specifically, our simulator is 6-DOF, includes flight condition estimation, a guidance system , a non-linear radome model, an actuator model that accounts for actuator bandwidth, fin limits, and fin rate limits, an inertial measurement unit (IMU) that models noise, bias, and placement distance from center of gravity, a radar receiver with line-of-sight (LOS) and range dependent noise, delays between IMU and radar measurements, modeling of aerodynamic uncertainty, and explicit modeling of LOS recovery using a strap down seeker. Importantly, we optimize and test the flight control systems taking into account aerodynamic model uncertainty, which would be expected in the hypersonic flight regime.

The simulator is shown in Fig. 1, where all systems run at 500 Hz. Modeling relative velocity estimation in this scenario has a significant performance impact [24]. Consequently, to better compare the effectiveness of the three flight control systems, we assume ground truth knowledge of relative velocity, and leave full relative state estimation to a separate paper [24]. However, we model flight condition estimation, including angle of attack, sideslip angle, Mach, altitude, and change in attitude from sensor measurements, as described later in Sections II.C.5 and II.C.7.

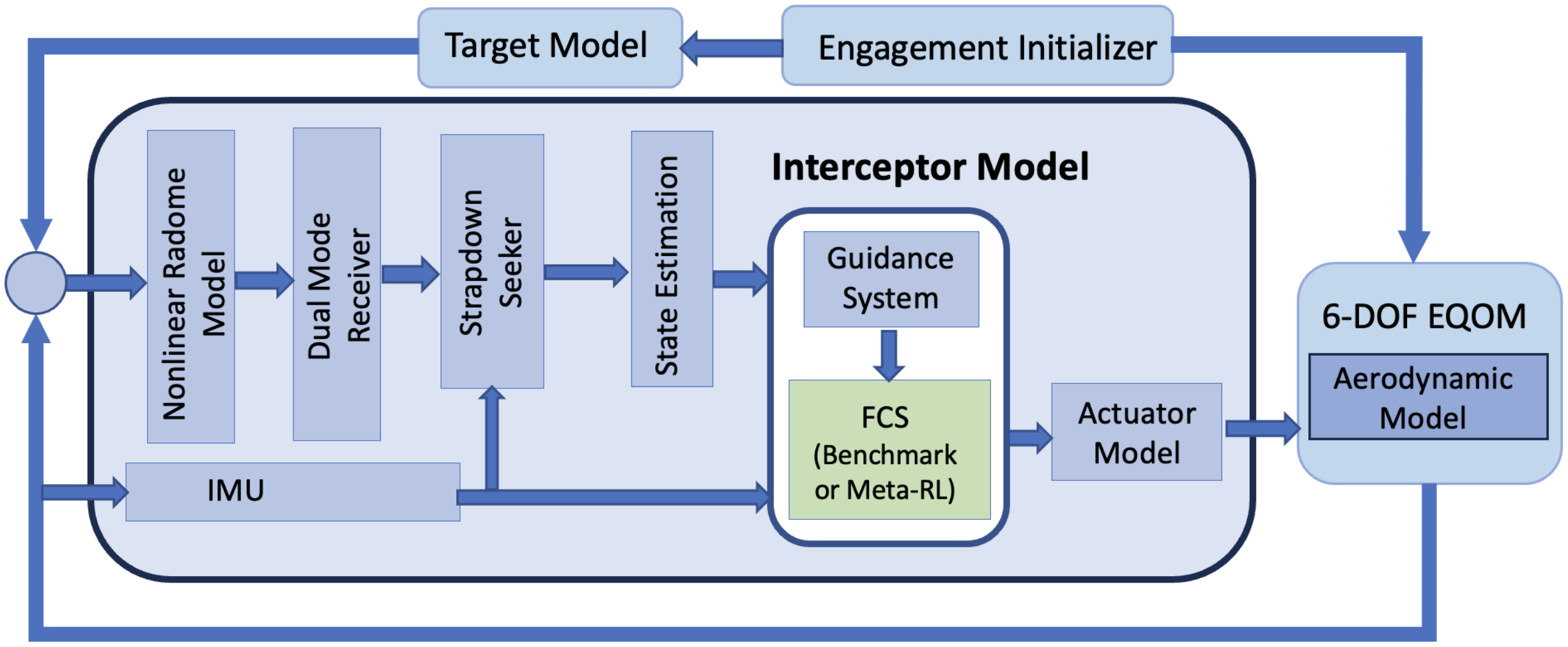

**Fig. 1 Simulator Diagram**

The interceptor G&C system must satisfy a load constraint, where the norm of the $y$ and $z$ body frame forces acting on the interceptor (see Section II.D) must be less than 75 g: $\|[F_y^{\text{B}}, F_z^{\text{B}}]\| < 75$ g, where g=9.81 m/s$^2$ is the magnitude of the gravitational acceleration at the Earth's surface. The look angle constraint requires the maximum angle between the body frame x-axis and the body frame LOS vector to be less than 45°. Rather than directly impose the look angle constraint by terminating an episode when the constraint is violated, we instead latch the G&C system observables at their value prior to the violation for as long as the constraint is violated. Consequently, it is possible to violate the constraint for short periods and maintain good performance, noting that this constraint is always violated close to the end of the engagement. Note that if we had included relative velocity estimation using a Kalman filter, the filter would propagate the state open loop (no measurement update) when the target is out of the field of view.

## A. Engagement Scenario

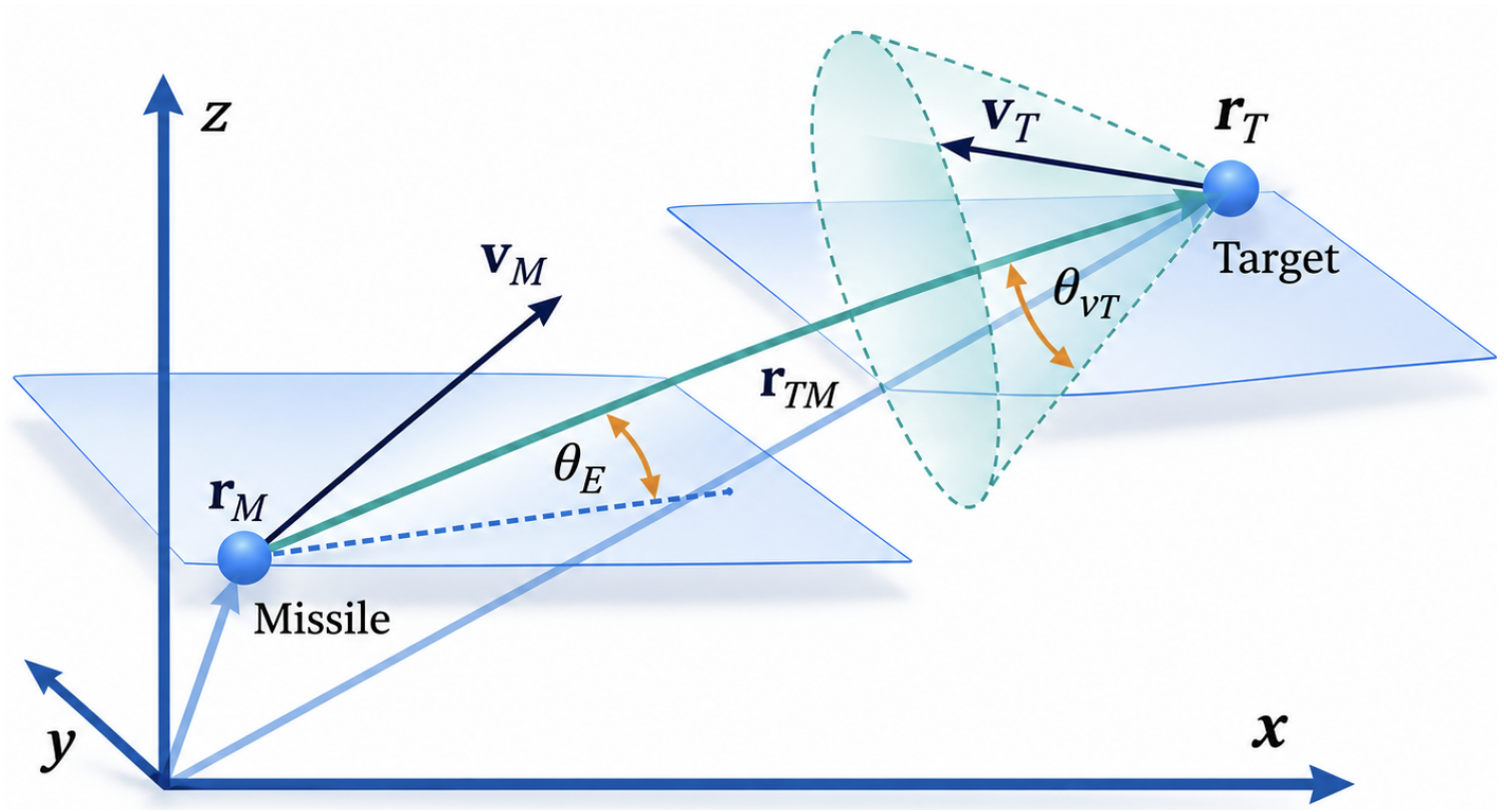

**Fig. 2 Engagement**

Referring to Fig. 2*, the missile position vector, missile velocity vector, target position vector, and target velocity vector are shown as $\mathbf{r}_\mathrm{M}$, $\mathbf{v}_\mathrm{M}$, $\mathbf{r}_\mathrm{T}$, $\mathbf{v}_\mathrm{T}$. We can also define the relative position and velocity vectors $\mathbf{r}_\mathrm{TM} = \mathbf{r}_\mathrm{T} - \mathbf{r}_\mathrm{M}$ and $\mathbf{v}_\mathrm{TM} = \mathbf{v}_\mathrm{T} - \mathbf{v}_\mathrm{M}$. The elevation angle $\theta_E$ is the angle between $\mathbf{r}_\mathrm{TM}$ and its projection onto the x-y plane. We randomly generate the target's initial velocity vector such that $\mathbf{v}_\mathrm{T}$ lies within a cone with axis $\mathbf{r}_\mathrm{TM}$ and half apex angle $\theta_{\mathbf{v}_\mathrm{T}}$. In $\mathbb{R}^3$ we define a heading error ($HE$) as the angle between the interceptor's initial velocity vector and the velocity vector associated with the lead angle required to put the interceptor on a collision heading with the target. The interceptor initial conditions are listed in Table 1.

**Table 1 Simulation Initial Conditions**

| Parameters Drawn Uniformly | Min | Max |
|---|---|---|
| Elevation Angle to Target $\theta_E$ (degrees) | 0 | 10 |
| Azimuth Angle to Target $\theta_A$ (degrees) | -180 | 180 |
| Altitude (m) | 15000 | 25000 |
| Missile Velocity Magnitude $\|\mathbf{v}_\mathrm{M}\|$ (m/s) | 2300 | 2600 |
| Heading Error (degrees) | 0 | 5 |
| Initial Angle of Attack $\alpha$ (degrees) | -5 | -5 |
| Initial Side Slip Angle $\beta$ (degrees) | -5 | 5 |
| Initial Range to Targeted Asset $\|\mathbf{r}_\mathrm{AT}\|$ (m) | 20000 | 25000 |
| Target Velocity Magnitude $\|\mathbf{v}_\mathrm{T}\|$ (m/s) | 3000 | 4000 |
| Target Velocity Cone Half Apex Angle $\theta_{\mathbf{v}_\mathrm{T}}$ (degrees) | 0 | 15 |
| Interceptor Maximum Acceleration (@ 2500 m/s 20000 m, applied to each component) (g) | - | 20 |
| Target Evasion Acceleration (@ 2500 m/s 20000 m, applied to each component) (g) | 0 | 4 |
| Target Weave Period (s) | 0.5 | 3.0 |

### B. Target Evasive Acceleration Model

During optimization, we consider three types of evasive maneuvers, a step maneuver, a weaving maneuver, and a jinking maneuver, and add different maneuver types to test the generalization capability of the optimized policies. The weaving maneuver has frequency shared for each axis and phase independent for each axis in $\mathbb{R}^3$, and randomized phase offset and frequency. In contrast, the jinking maneuver is only randomized in frequency, shared for each axis. Each of these maneuvers has a 1/3 probability of being selected at the start of each maneuver. At each simulation step, the target's evasive acceleration vector is modified by removing the component parallel with the target velocity vector, and then adjusted for dynamic pressure, with a nominal acceleration capability of 4 g at an altitude of 20 km and a speed of 2500 m/s. Thus, the target acceleration vector is always orthogonal to the target velocity. The maneuvers have varying acceleration levels and random start time, duration, and switching time. At the start of each episode, with probability 0.5 the maneuvers in that episode use the full evasion acceleration capability, and with probability 0.5 the full acceleration is scaled uniformly by a factor uniformly distributed between 0 and 1. Note that the target drag orthogonal to the interceptor LOS is also perceived as a maneuver; the target drag model is shown in Section II.E Sample target maneuvers are shown in Fig. 3, which show evasive acceleration only (not drag orthogonal to LOS).

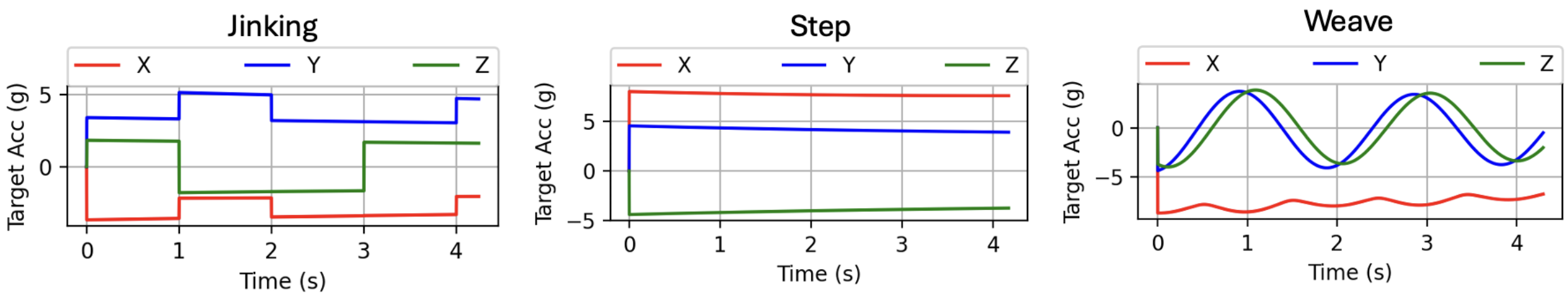

**Fig. 3 Sample Target Maneuvers (only evasive acceleration shown**

*In this figure, the illustrated vectors are not in general within the x-z plane

## C. Interceptor Components

*1. Radome Model*

We calculate the refracted azimuth $\tilde{\theta}_u$ and elevation $\tilde{\theta}_v$ angles as shown in Eqs. (1a) through (1b), where our nominal value of $A$ is 0.5 mrad, and $K_u$, $K_v$ are set to 3.0. Note we assume that there is a combination of software (such as compensation lookup tables) and hardware (such as radome shaping) that reduce the effective radome refraction to these values.

$$\tilde{\theta}_u = \theta_u + A\left(0.75\frac{\theta_u}{\pi/2} + 0.25\cos\left(\frac{2\pi}{k_u}\theta_u\right)\right) \tag{1a}$$

$$\tilde{\theta}_v = \theta_v + A\left(0.75\frac{\theta_v}{\pi/2} + 0.25\cos\left(\frac{2\pi}{k_v}\theta_v\right)\right) \tag{1b}$$

The refraction angle $\theta_R$ is shown in Fig. 4 for various values of $k$. Note that the radome slope $\dfrac{\partial\theta_R}{\partial\theta_L}$ is given by the slope of the curves in the figure.

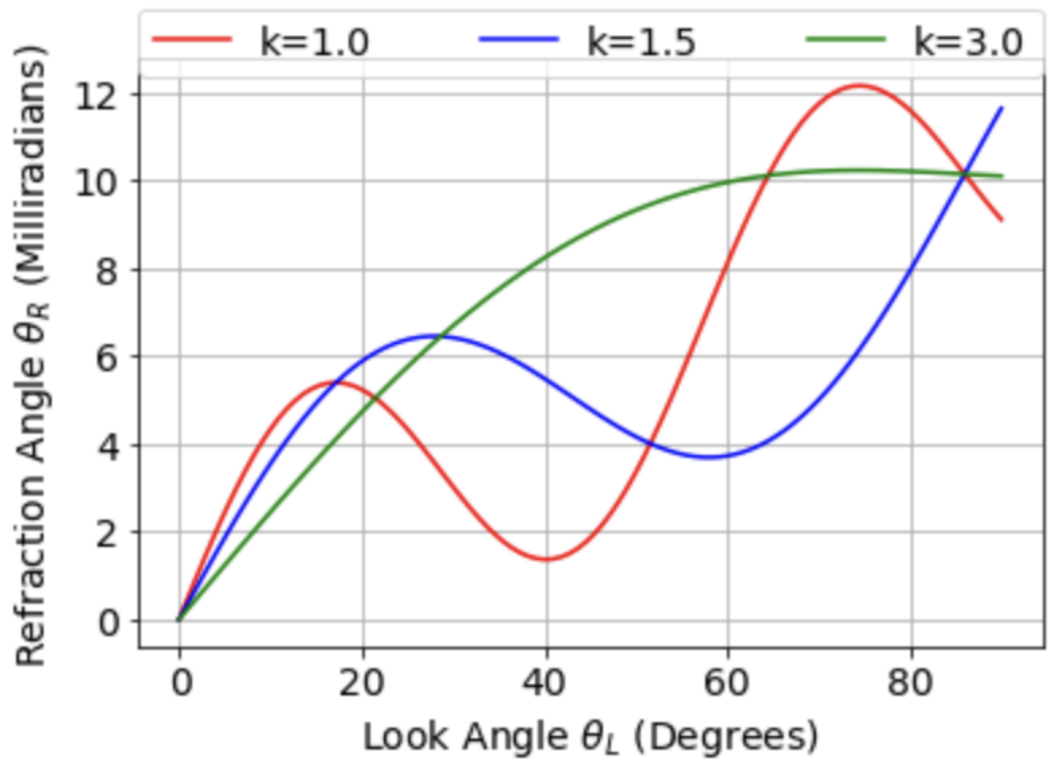

**Fig. 4 Radome Refraction Angle as Function of Look Angle**

*2. Seeker Model*

We model sensor noise (elevation, azimuth, and range) as Gaussian noise, using a standard deviation of 1% of range for range dependent range noise and 0.2 mrad for angle noise. Obtaining this level of angle noise requires a dual model seeker (radar for range, infrared for angles). This improves angular resolution, target discrimination, and resistance to countermeasures. Note that our assumed noise level is post signal processing. We model the lag of the seeker signal processor as a pure delay of 4 ms.

*3. Rate Gyro Model*

Our rate gyro model corrupts the ground truth rotational velocity $\boldsymbol{\omega}^{\mathrm{B}}$ with both scale factor bias and Gaussian noise, as shown in Eq. (2a), where $\mathcal{U}(a, b, n)$ denotes an $n$ dimensional uniformly distributed random variable bounded by $(a, b)$ that is generated at the start of each episode, and each dimension of the random variable is independent. We use $\epsilon_\omega = 0.001$ and $\sigma_\omega = 0.8°$. The rotational velocity is passed through a two pole low pass filter with a 100 Hz natural frequency and 0.707 damping, as shown in Eq. 2b.

$$\boldsymbol{\omega}_1^{\mathrm{B}} = \boldsymbol{\omega}(1 + \mathcal{U}(-\epsilon_\omega, \epsilon_\omega, 3)) \tag{2a}$$

$$\boldsymbol{\omega}_{\mathrm{obs}}^{\mathrm{B}} = \mathrm{LPF}(\boldsymbol{\omega}_1^{\mathrm{B}}) + \mathcal{N}(0, \sigma_\omega, 3) \tag{2b}$$

*4. Accelerometer Model*

We account for accelerometer placement using Eq. (3a), with $\mathbf{c} = \mathbf{r}_{cg} - \mathbf{r}_{acc}$, where $\mathbf{r}_{acc}$ is the accelerometer location, which we assume to be -25 cm using the reference frame shown in Fig. 5. Similarly to the rate gyro model, the body frame acceleration is corrupted with a scale factor error as shown in Eqs. (3b) and (3c), where we use $\boldsymbol{\epsilon}_{acc} = 0.001$, and "LPF" models the accelerometer dynamics as a two pole low pass filter with a 100 Hz natural frequency and 0.707 damping, and $\sigma_{acc} = 0.1$ m/s$^2$.

$$\mathbf{a}_1^{\mathrm{B}} = \mathbf{a}^{\mathrm{B}} + [\dot{\omega}\times]\mathbf{c} + [\omega\times][\omega\times](\mathbf{c} + \delta\mathbf{c}) \tag{3a}$$

$$\mathbf{a}_2^{\mathrm{B}} = \mathrm{LPF}(\mathbf{a}_1^{\mathrm{B}}) \tag{3b}$$

$$\mathbf{a}_{obs}^{\mathrm{B}} = \mathbf{a}_2^{\mathrm{B}}(1 + \mathcal{U}(-\boldsymbol{\epsilon}_{acc}, \boldsymbol{\epsilon}_{acc}, 3)) + \mathcal{N}(0, \sigma_{acc}, 3) \tag{3c}$$

*5. Line of Sight Recovery*

Referring to Fig. 1, the refracted body frame line of sight unit vector $\hat{\mathbf{r}}_{\mathrm{TM}}^{\mathrm{B}}$ from the dual-mode receiver block is input to the LOS recovery block. This block uses the estimated change of attitude since the start of the engagement $\mathbf{dq}_{obs}$ to rotate $\hat{\mathbf{r}}_{\mathrm{TM}}^{\mathrm{B}}$ into the reference frame corresponding to the interceptor's attitude at the start of the engagement. Specifically, at each guidance step we map $\mathbf{dq}_{obs}$ to a direction cosine matrix $\mathbf{C}(\mathbf{dq}_{obs})$, and compute the recovered (stabilized) LOS as $\hat{\mathbf{r}}_{\mathrm{TM}}^{\mathrm{S}} = [\mathbf{C}(\mathbf{dq}_{obs})]^{\mathrm{T}}\hat{\mathbf{r}}_{\mathrm{TM}}^{\mathrm{B}}$.

The estimated change in attitude $\mathbf{dq}_{obs}$ is parameterized as a quaternion, and is computed by integrating $\boldsymbol{\omega}$ as shown in Equation 4, where $\mathbf{dq}_{obs}$ is reset at the start of each episode to $[1, 0, 0, 0]$. In our simulation model, we approximate this integration using fourth-order Runge-Kutta integration. In the unrealistic case where we have access to the ground truth value of $\boldsymbol{\omega}$, this completely decouples $\hat{\mathbf{r}}_{\mathrm{TM}}$ from body rotation. However, since $\boldsymbol{\omega}_{obs}$ is corrupted with a scale factor error and Gaussian noise, in general $\mathbf{dq}_{obs} \neq \mathbf{dq}$, and the LOS recovery will be imperfect. The combination of imperfect LOS recovery and radome refraction result in a false indication of target motion. This results in increased miss distance, and can potentially destabilize the G&C system. This is discussed in more detail in [6, 7, 25].

$$\begin{bmatrix} \dot{\mathrm{dq}}_0 \\ \dot{\mathrm{dq}}_1 \\ \dot{\mathrm{dq}}_2 \\ \dot{\mathrm{dq}}_3 \end{bmatrix} = \frac{1}{2} \begin{bmatrix} \mathrm{dq}_0 & -\mathrm{dq}_1 & -\mathrm{dq}_2 & -\mathrm{dq}_3 \\ \mathrm{dq}_1 & \mathrm{dq}_0 & -\mathrm{dq}_3 & \mathrm{dq}_2 \\ \mathrm{dq}_2 & \mathrm{dq}_3 & \mathrm{dq}_0 & -\mathrm{dq}_1 \\ \mathrm{dq}3 & -\mathrm{dq}_2 & \mathrm{dq}_1 & \mathrm{dq}_0 \end{bmatrix} \begin{bmatrix} 0 \\ \omega_{obs_0} \\ \omega_{obs_1} \\ \omega_{obs_2} \end{bmatrix} \tag{4}$$

*6. Guidance System*

The guidance system implements true proportional navigation (TPN) [26] to calculate a commanded acceleration $\mathbf{a}_{com}^{\mathrm{S}}$ from the LOS rotation rate, as shown in Eqs. (5a) through (5c), where $v_c = -\frac{\mathbf{r}_{\mathrm{TM}}^{\mathrm{S}} \cdot \mathbf{r}_{\mathrm{TM}}^{\mathrm{S}}}{\|\mathbf{r}_{\mathrm{TM}}^{\mathrm{S}}\|}$. This commanded acceleration is then rotated into the body frame using the estimated change in attitude $\mathbf{dq}$ and adjusted using g-limits. These g-limits assume a maximum acceleration of 25 g at an altitude of 20 km and a speed of 2500 m/s, and are adjusted for dynamic pressure. This body frame commanded commanded acceleration is then used to compute the acceleration tracking error according to $\mathbf{a}_{err}^{\mathrm{B}} = \mathbf{a}_{com}[y, z]^{\mathrm{B}} - \mathbf{a}_{obs}[y, z]^{\mathrm{B}}$, where $\mathbf{a}_{obs}^{\mathrm{B}}$ is the accelerometer measurement and $[y, z]$ indicates we are computing the error over the body frame $y$ and $z$ components. We use $N = 4$ for the navigation ratio.

$$\boldsymbol{\Omega} = \frac{\mathbf{r}_{\mathrm{TM}}^{\mathrm{S}} \times \mathbf{v}_{\mathrm{TM}}^{\mathrm{S}}}{\mathbf{r}_{\mathrm{TM}}^{\mathrm{S}} \cdot \mathbf{r}_{\mathrm{TM}}^{\mathrm{S}}} \tag{5a}$$

$$\mathbf{a}_{com}^{\mathrm{S}} = -N v_c \left( \frac{\mathbf{r}_{\mathrm{TM}}^{\mathrm{S}}}{\|\mathbf{r}_{\mathrm{TM}}^{\mathrm{S}}\|} \times \boldsymbol{\Omega} \right) \tag{5b}$$

$$\mathbf{a}_{com}^{\mathrm{B}} = [\mathbf{C}(\mathbf{dq}_{obs})]\mathbf{a}_{com}^{\mathrm{S}} \tag{5c}$$

*7. Flight Condition Estimation*

The FCS requires an estimate of angle of attack $\alpha$, sideslip angle $\beta$, altitude $h$, and Mach $M$. These observations are generated by integrating $\mathbf{a}_{obs}^{\mathrm{B}}$ to obtain an estimate of body frame velocity and position, allowing computation of estimated interceptor velocity $\mathbf{v}^{\mathrm{B}}_{\mathrm{obs}}$ and position $\mathbf{r}^{\mathrm{B}}_{\mathrm{obs}}$, with observed angle of attack $\alpha_{\mathrm{obs}}$ and sideslip angle $\beta_{\mathrm{obs}}$ computed from $\mathbf{v}^{\mathrm{B}}_{\mathrm{obs}}$. Estimated body frame position $\mathbf{r}^{\mathrm{B}}_{\mathrm{obs}}$ is rotated into the inertial frame using estimated attitude, allowing computation of altitude $h_{\mathrm{obs}}$. Mach $M$ is computed from estimated altitude and speed using an interpolated Earth atmospheric model table. Finally, for roll stabilization, the FCS requires an estimate of roll ($\theta_{\mathrm{roll_{obs}}}$), which is computed by integrating $\omega_{\mathrm{obs}_x}$.

#### *8. Actuator Model*

The channel commanded deflections $\boldsymbol{\delta}_{\mathrm{C}} = [\delta_{\mathrm{P}}, \delta_{\mathrm{Y}}, \delta_{\mathrm{A}}, \delta_{\mathrm{SB}}]$ (pitch, yaw, roll, and speed brake channel commands) from the FCS are converted to physical fin deflections $\boldsymbol{\delta}_{\mathrm{F}} = [\delta_1, \delta_2, \delta_3, \delta_4]$ using the fin mix matrix shown in Eq. (6) and the transformations shown in Eqs. (7a) and (7b). Note that the speed brake is unused.

$$\mathbf{M}_{\mathrm{fin}} = \begin{bmatrix} 1 & 0 & 1 & -1 \\ 0 & 1 & 1 & 1 \\ -1 & 0 & 1 & -1 \\ 0 & -1 & 1 & 1 \end{bmatrix} \tag{6}$$

$$\boldsymbol{\delta}_{\mathrm{C}} = \mathbf{M}_{\mathrm{fin}}^{-1}\boldsymbol{\delta}_{\mathrm{F}} \tag{7a}$$

$$\boldsymbol{\delta}_{\mathrm{F}} = \mathbf{M}_{\mathrm{fin}}\boldsymbol{\delta}_{\mathrm{C}} \tag{7b}$$

$\boldsymbol{\delta}_{\mathrm{F}}$ is then clipped to remain within the maximum deflection of 30°, and the slew rate is limited to 500°/ s. Each element of $\boldsymbol{\delta}_{\mathrm{F}}$ is then passed through second order actuator dynamics corresponding to the transfer function shown in Eq. (8), where we use $\zeta_{\mathrm{ACT}} = 0.7$ and $\omega_{\mathrm{ACT}} = 314$ rad/s (50Hz). The output of the actuator dynamics is then referred to as $\boldsymbol{\delta}_{\mathrm{F}}$ in the remainder of the document.

$$\frac{1}{1 + \frac{2\zeta_{\mathrm{ACT}}}{\omega_{\mathrm{ACT}}}s + \frac{s^2}{\omega^2_{\mathrm{ACT}}}} \tag{8}$$

### D. Aerodynamic Model

The interceptor geometry is taken from [27] and is shown in Fig. 5, with the nominal center of gravity set to -0.54 m. Our nominal aerodynamic model is derived from slender body and Newtonian impact theory [28, 29]. The method in [29] is extended to 6-DOF using a method similar to that in [22], but with a higher fidelity axial force model, rolling moment model, and the addition of a cross channel interference term for control effectiveness. The vehicle diameter is $d = 0.105$ m. The length is $x_{\mathrm{L}} = 1.05$ m, and the inertia tensor is given in Figure 9. The vehicle mass is 13.5 kg, the reference length is $\mathrm{L_{ref}} = d$, and the reference area is $\mathrm{S_{ref}} = \pi(d/2)^2$ . The open loop response to flight control commanded deflections are shown in Figs. 6 and 7, where $\delta_{\mathrm{C}} = \delta_{\mathrm{P}}, \delta_{\mathrm{Y}}, \delta_{\mathrm{A}}$ are the FCS commanded channel deflections as described in Section II.C.8. We see that the vehicle is slightly open-loop unstable.

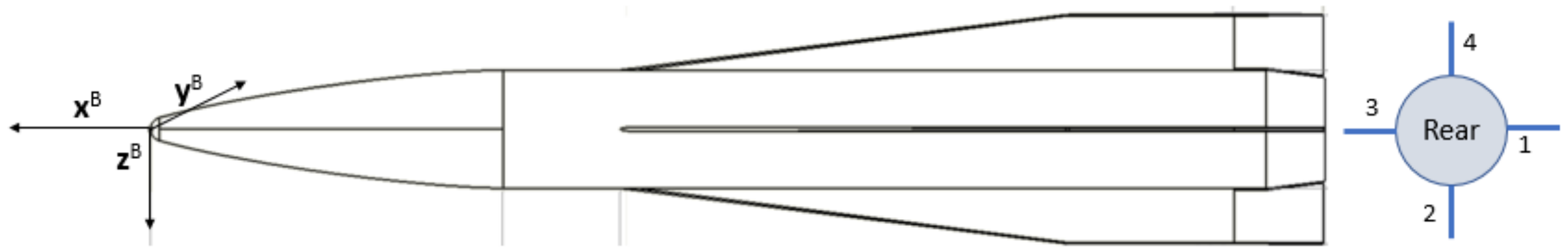

**Fig. 5 Interceptor Geometry**

$$\mathbf{J} = \begin{bmatrix} 0.02 & 0 & 0 \\ 0 & 0.95 & 0 \\ 0 & 0 & 0.95 \end{bmatrix} \tag{9}$$

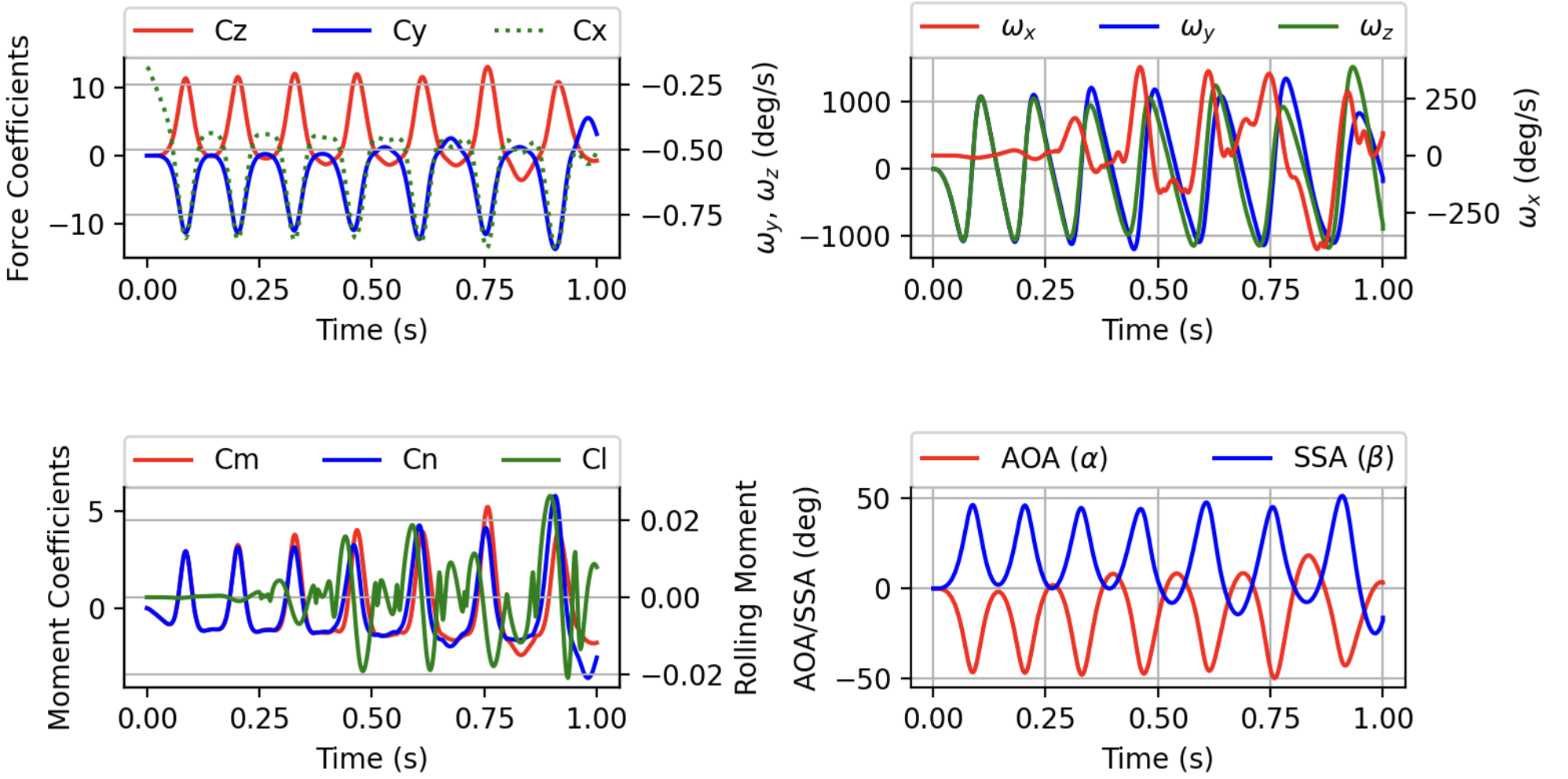

**Fig. 6 Open Loop Airframe Response $\delta_C = [30, 30, 0]$, Altitude=15 km, speed=2500 m/s**

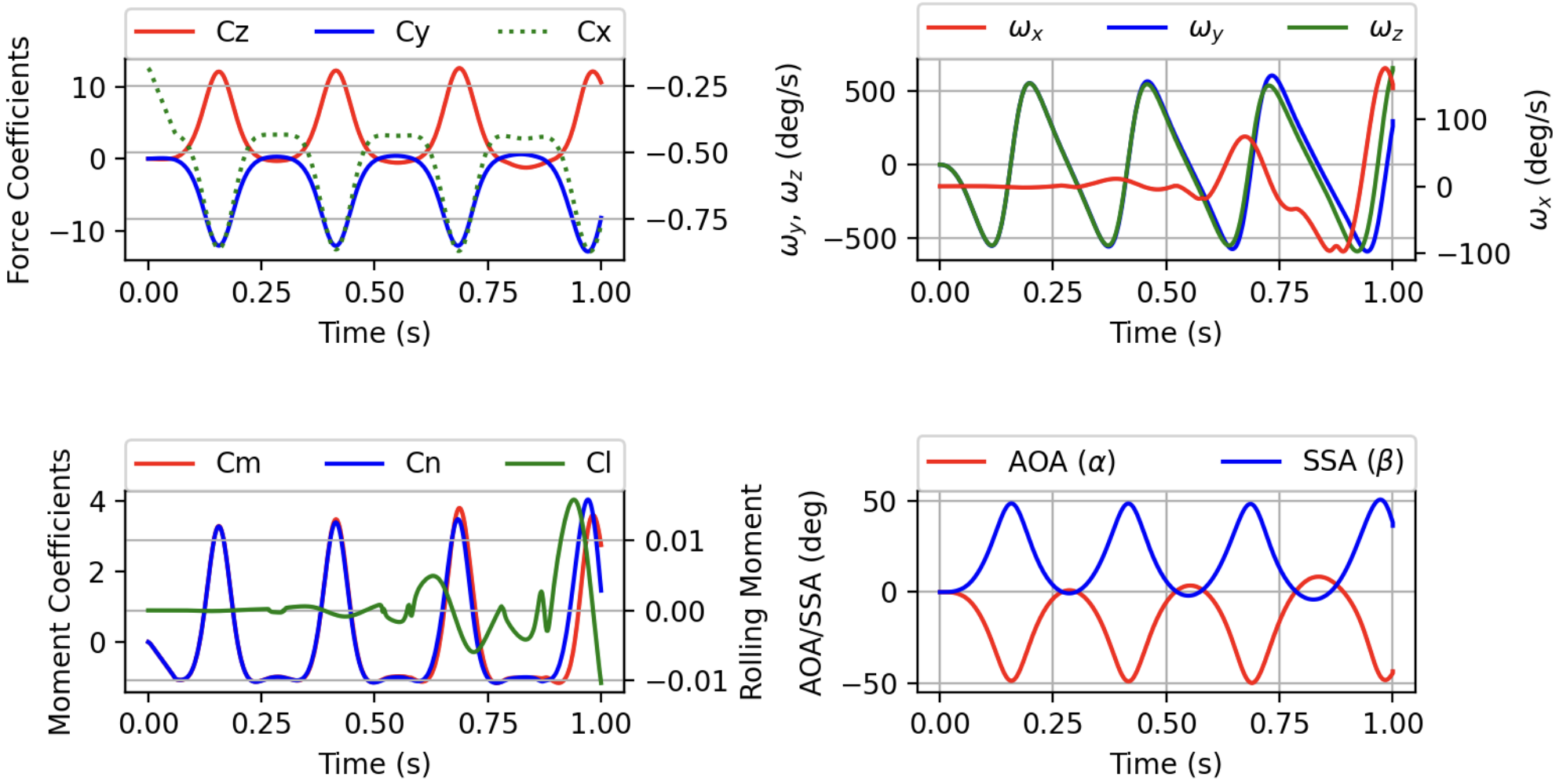

**Fig. 7 Open Loop Airframe Response $\delta_C = [30, 30, 0]$, Altitude=25 km, speed=2500 m/s**

We model aerodynamic uncertainty by randomly modifying the shape of the aerodynamic coefficient curves for each simulated trajectory. To do so, we first bias the coefficients returned from the nominal aerodynamic model as follows. At the start of each episode with probability $p = 0.5$ we uniformly sample an aerodynamic coefficient bias $\boldsymbol{\epsilon}_{\text{coeff}} \in \mathbb{R}^4$ between some bounds $+/-\epsilon$, otherwise a coefficient bias is (independently) set to either $-\epsilon$ or $\epsilon$ with equal probability, noting that in both cases the bias applied to the coefficient groups $(C_z, C_y)$ ,$(C_m, C_n)$, $C_x$, and $C_l$ are independent. We assume that the bias on $C_z$ and $C_y$ is identical to symmetry (and similarly for $C_m$ and $C_n$). Then, for the remainder of the episode, the force and moment coefficients are biased using the sampled value of $\boldsymbol{\epsilon}_{\text{coeff}}$. We account for CP/CG shifting by modifying the moments as shown in Eqs. (10a) through (10d). Here $\boldsymbol{\Delta}_{\text{CP}}$ is uniformly generated at the start of each episode and MRP is the moment reference point, which in this work is set to the CG. Due to symmetry, errors in $\Delta_{\text{CP}_y}$ and $\Delta_{\text{CP}_z}$ will mostly cancel.

$$\Delta_{\text{CGCP}} = \frac{(\text{MRP} - \text{CG}) + \Delta_{\text{CP}}}{L_{\text{ref}}} \times [Cx, Cy, Cz] \tag{10a}$$

$$Cl_{\text{adj}} = Cl + \Delta_{\text{CGCP}_x} \tag{10b}$$

$$Cm_{\text{adj}} = Cm + \Delta_{\text{CGCP}_y} \tag{10c}$$

$$Cn_{\text{adj}} = Cn + \Delta_{\text{CGCP}_z} \tag{10d}$$

### E. Equations of Motion

Let $\mathbf{v}^{\text{B}} \equiv [u, v, w]$ denote the body frame interceptor velocity. The angle of attack $\alpha$ and sideslip angle $\beta$ are then computed as shown in Eqs. (11a) through (11b).

$$\alpha = \arctan \frac{w}{u} \tag{11a}$$

$$\beta = \arctan \frac{v}{u} \tag{11b}$$

$$\tag{11c}$$

The atmospheric density $\rho$ and Mach number $M$ are calculated using the 1976 standard Earth atmospheric temperature model. The interceptor control surface deflections $\boldsymbol{\theta}_{\text{ctrl}} = [\theta_{\text{LT}}, \theta_{\text{RT}}, \theta_{\text{BT}}, \theta_{\text{UT}}]$, dynamic pressure $q_o = \frac{1}{2}\rho V^2$ with $V = \|\mathbf{v}^B\|$, rotational velocity vector $\boldsymbol{\omega}$, $M$, $\alpha$ and $\beta$ are then input to the aerodynamic model described in Section II.D to compute the body frame force $\mathbf{F}^{\text{B}}$ and torque $\mathbf{L}^{\text{B}}$.

The rotational velocity $\boldsymbol{\omega}$ is updated by integrating the Euler rotational equations of motion, as shown in Equation (12), where $\mathbf{J}$ is the interceptor's inertia tensor, and with the skew symmetric operator $[\mathbf{a}\times]$ defined in Eq. (13).

$$\mathbf{J}\dot{\boldsymbol{\omega}} = -[\boldsymbol{\omega}\times]\mathbf{J}\boldsymbol{\omega} + \mathbf{L}^{\text{B}} \tag{12}$$

$$[\mathbf{a}\times] \equiv \begin{bmatrix} 0 & -a_3 & a_2 \\ a_3 & 0 & -a_1 \\ -a_2 & a_1 & 0 \end{bmatrix} \tag{13}$$

The interceptor's attitude $\mathbf{q}$ is updated by integrating the differential kinematic equations shown in Equation (14), where the interceptor's attitude is parameterized using the quaternion representation and $\boldsymbol{\omega}_i$ denotes the $i^{th}$ component of the rotational velocity vector $\boldsymbol{\omega}_B$.

$$\begin{bmatrix} \dot{q}_0 \\ \dot{q}_1 \\ \dot{q}_2 \\ \dot{q}_3 \end{bmatrix} = \frac{1}{2} \begin{bmatrix} q_0 & -q_1 & -q_2 & -q_3 \\ q_1 & q_0 & -q_3 & q_2 \\ q_2 & q_3 & q_0 & -q_1 \\ q_3 & -q_2 & q_1 & q_0 \end{bmatrix} \begin{bmatrix} 0 \\ \omega_0 \\ \omega_1 \\ \omega_2 \end{bmatrix} \tag{14}$$

The interceptor body frame velocity $\mathbf{v}^{\text{B}}$ is updated by integrating the differential equation given in Eq. 15, where $m$ is the interceptor mass, $\mathbf{C}_{\text{BN}}(\mathbf{q})$ is the DCM mapping from the inertial to body frame given the interceptor attitude $\mathbf{q}$, and $g = -9.81$ m/s$^2$.

$$\dot{\mathbf{v}}^{\text{B}} = -[\boldsymbol{\omega}\times]\mathbf{v}^{\text{B}} + \frac{\mathbf{F}^{\text{B}}}{m} + [\mathbf{C}_{\text{BN}}(\mathbf{q})]^T \begin{bmatrix} 0 & 0 & g \end{bmatrix} \tag{15}$$

$\mathbf{v}^{\text{B}}$ is then rotated into the North-East-Down (NED) inertial frame as shown in Eq. (16a), and $\mathbf{v}_{\text{M}}$ is calculated by rotating the vector into the East-North-Up (ENU) reference frame depicted in Fig. 2.

$$\mathbf{v}^{\text{N}} = \mathbf{C}_{\text{BN}}(\mathbf{q})\mathbf{v}^{\text{B}} \tag{16a}$$

$$\mathbf{v}_{\text{M}} = \mathbf{C}_{\text{NE}}\mathbf{v}^{\text{N}} \tag{16b}$$

The interceptor inertial frame position $\mathbf{r}_{\mathrm{M}}$ is then updated by integrating Eq. (17)

$$\dot{\mathbf{r}}_{\mathrm{M}} = \mathbf{v}_{\mathrm{M}} \tag{17}$$

The target inertial frame position $\mathbf{r}_{\mathrm{T}}$ and velocity unadjusted for drag $\tilde{\mathbf{v}}_{\mathrm{T}}$ are then updated by integrating Eqs. (18a) and (18b). The target speed $V_{\mathrm{T}} = \|\mathbf{v}_{\mathrm{T}}\|$ is adjusted for drag by integrating Eq. (18c), where at the start of each episode we randomly generate $k_{\mathrm{T}}$ from 0.1 to 0.25 and $\mathrm{cd}_0$ from 0.05 to 0.10. We use $m_{\mathrm{T}} = 450$ kg.

$$\dot{\mathbf{r}}_{\mathrm{T}} = \mathbf{v}_{\mathrm{T}} \tag{18a}$$

$$\dot{\tilde{\mathbf{v}}}_{\mathrm{T}} = \mathbf{a}_{\mathrm{T}} \tag{18b}$$

$$\dot{V}_{\mathrm{T}} = -\frac{\rho(h_{\mathrm{T}})V_{\mathrm{T}}^2\mathrm{cd}_0}{2m_{\mathrm{T}}} - k_{\mathrm{T}}\|[\mathbf{a}_{\mathrm{T}}\|] \tag{18c}$$

The target velocity is then computed as shown in Eq. (19)

$$\mathbf{v}_{\mathrm{T}} = V_{\mathrm{T}}\frac{\tilde{\mathbf{v}}_{\mathrm{T}}}{\|\tilde{\mathbf{v}}_{\mathrm{T}}\|} \tag{19}$$

The equations of motion are updated using fourth order Runge-Kutta integration. For ranges greater than 50 m, a timestep of 5 ms is used, and for the final 50 m of homing, a timestep of 25 us is used in order to more accurately (within 0.125 m) measure miss distance; this technique is borrowed from [30].

## III. 3-Loop Autopilot FCS Benchmark

The benchmark FCS (FCS-3L) for this application uses a three-loop autopilot for pitch and yaw control. We assume perfect roll control for the benchmark (and for the augmented benchmark, the FCS-3L-RL), obtained by forcing the x-component of rotational velocity to zero, making this a more difficult benchmark to beat. The three-loop autopilot design and gain scheduling are taken from [9], with the modification for an unstable airframe (the linearized airframe is unstable) taken from [31], and modified for our vehicle geometry. This gain scheduling was suboptimal for altitudes above 15 km. Therefore, we performed a series of grid searches on the open-loop crossover frequency $\omega_{\mathrm{CR}}$, damping $\zeta$, and time constant $\tau$ at three design points: 2500 m/s 15 km, 2500 m/s 20 km, and 2500 m/s 25 km. For a given design point, each design (a tuple of $\omega_{\mathrm{CR}}, \zeta, \tau$) was simulated for 5000 episodes with up to 10% aerodynamic coefficient bias and 1% center of pressure variation (see Section II.D). We considered a design stable if it resulted in no load constraint violations. We chose this approach as traditional stability metrics such as phase margin with a linearized system can be overly conservative, as a large gain margin can be required to compensate for differences between the linearized and actual system. This approach also gives a better comparison to the FCS-RL, which being inherently nonlinear, cannot be analyzed using linear system metrics. For each FCS parameter tuple at a given design point, we chose the best performing set of parameters that resulted in a stable design. During FCS operation, we interpolate between the parameters at each of the three design points using linear interpolation. This approach is different than what is typically used for interceptor flight control systems, where the gains are optimized at a given design point using step response metrics, but has the advantage that the gains can be optimized considering the dynamics of both the guidance and flight control systems, which is more directly comparable to the approach used for FCS-RL optimization. A block diagram of the pitch and yaw channels for the FCS-3L is illustrated in Fig. 8, where $a_{\mathrm{com}}, a_{\mathrm{obs}}, \omega_{\mathrm{obs}}$ coming from either the pitch or yaw channel from the guidance system, accelerometer, and rate gyro, respectively, and $\delta_{\mathrm{com}}$ being either the pitch or yaw commanded channel deflection. The relationship between the block diagram gains and $\omega_{\mathrm{CR}}, \zeta, \tau$ can be found in [9]. Step response plots for this benchmark can be found in Section VI.A, Fig. 13. We did not add adaptive control, as previous work [32] showed that MRAC was ineffective in modeling this intercept scenario due to the short terminal phase homing time (which is unavoidable with high closing speeds and a realistic seeker range) and the other factors discussed in Section I .

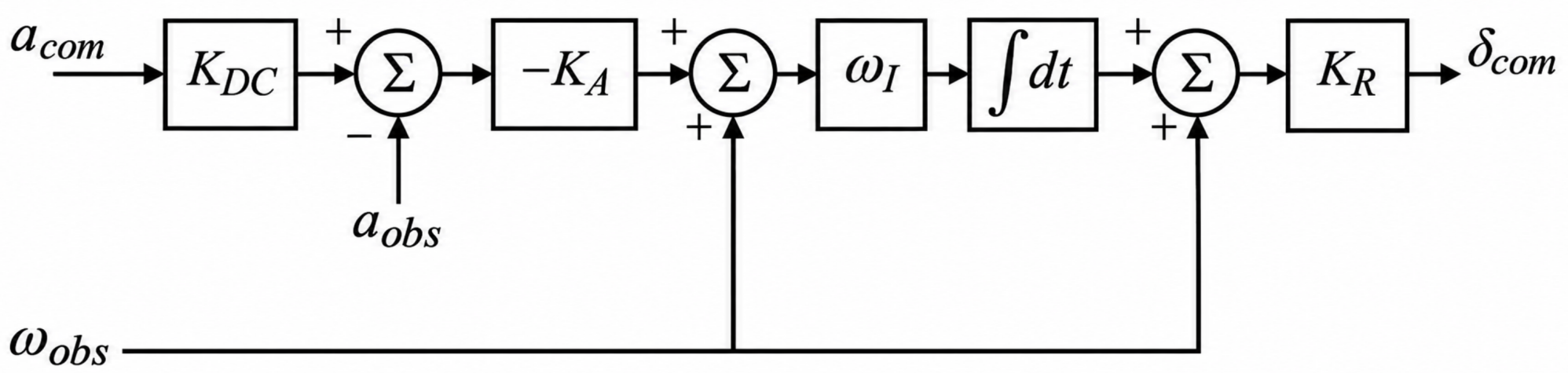

**Fig. 8 Three-Loop Autopilot Block Diagram**

# IV. Meta-RL Optimization Framework

Note that in this section, unless a variable has the "B" (body frame) superscript, it is assumed to be in the LOS stabilized ("S") reference frame.

## A. Overview of Reinforcement Learning

In the reinforcement learning framework, an agent learns through episodic interaction with an environment how to successfully complete a task using a policy $\pi$ that maps observations $\mathbf{o}$ to actions $\mathbf{u}$, i.e., $\mathbf{u} = \pi(\mathbf{o})$. In actor-critic RL algorithms, policy optimization makes use of a value function $v = V^{\pi}(\mathbf{x})$ that estimates the estimated value of being in state $\mathbf{x}$ while following policy $\pi$. Note that the value function can take either ground truth states $\mathbf{x}$, observations $\mathbf{o}$, or some combination of both, as only the policy is deployed. The environment initializes an episode by generating a ground truth state $\mathbf{x}$, mapping this state to an observation, and passing the observation to the agent. The agent uses this observation to generate an action that is sent to the environment; the environment then uses the action and the current ground truth state to generate the next state and a scalar reward signal $r(\mathbf{x}, \mathbf{u})$. The reward and the observation corresponding to the next state are then passed to the agent. The process repeats until the environment terminates the episode, with the termination signaled to the agent via a done signal. In this work, an episode terminates when the closing velocity $v_c$ turns negative, or a path constraint is violated. Trajectories collected over a set of episodes (referred to as rollouts) are collected during interaction between the agent and environment, and used to update the policy and value functions. The interface between agent and environment is depicted in Fig. 9, where the environment instantiates the simulator and the system model shown in Fig. 1.

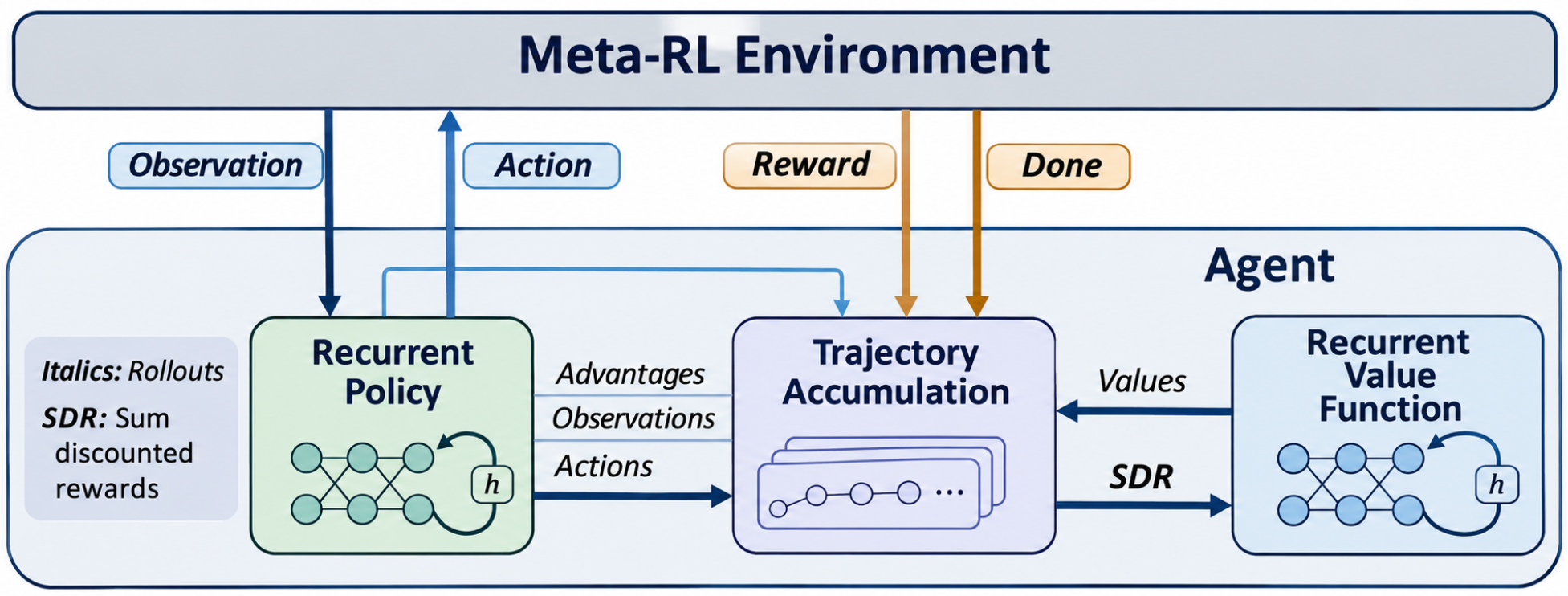

**Fig. 9 Environment-Agent Interface**

Meta-RL differs from generic reinforcement learning in that the agent learns over an ensemble of environments. For example, each environment can contain a unique aerodynamic model, target maneuver model, actuator model, and other peripheral components. Similar to [33], we implement both the policy and value function as a deep recurrent neural network, with each network having a hidden state $\mathbf{h}$ that is a function of the history of observations $\mathbf{o}$. Thus, the policy and value function become $\mathbf{u} = \pi_{\theta}(\mathbf{o}, \mathbf{h})$ and $v = V^{\pi}(\mathbf{x}, \mathbf{h})$. We optimize the policy using a customized version of the proximal policy optimization algorithm (PPO) [34]. Maximizing the PPO objective function requires learning hidden layer parameters $\boldsymbol{\theta}_\mathrm{h}$ that result in $\mathbf{h}$ evolving in response to the history of $\mathbf{o}$ and $\mathbf{u}$ in a manner that facilitates fast adaptation to environments encountered during optimization, as well as novel environments outside of the training distribution. The optimized policy then adapts real-time to conditions experienced during deployment. Importantly, the network parameters remain fixed during deployment with adaptation occurring through the evolution of $\mathbf{h}$.

The network architectures used in this work are shown in Table 2, where $n_\mathrm{hi}$ is the number of units in layer $i$, obs_dim is the observation dimension, and act_dim is the action dimension. The policy and value functions are periodically updated during optimization after accumulating trajectory rollouts of 30 simulated episodes.

**Table 2 Policy and Value Function network architecture**

| | Policy Network | | Value Network | |
|---|---|---|---|---|
| Layer | # units | activation | # units | activation |
| hidden 1 | 10∗obs_dim | tanh | 10∗obs_dim | tanh |
| hidden 2 | $\sqrt{n_\mathrm{h1} * n_\mathrm{h3}}$ | tanh | $\sqrt{n_\mathrm{h1} * n_\mathrm{h3}}$ | tanh |
| hidden 3 | 10 ∗ act_dim | tanh | 5 | tanh |
| output | act_dim | linear | 1 | linear |

# V. Meta-RL Problem Formulation for Gain Adaptation (FCS-3L-RL)

The agent observation $\mathbf{o}$ is shown in Eqs. (20), where $\alpha_\mathrm{obs}$, $\beta_\mathrm{obs}$, $M_\mathrm{obs}$, $h_\mathrm{obs}$, $\boldsymbol{\omega}_\mathrm{obs}$, and $\mathbf{a}_\mathrm{err}$, are generated as shown Sections II.C.3), II.C.4, and II.C.6. $\boldsymbol{\delta}_\mathrm{C}$ are the commanded channel deflections from Section II.C.8 and derived from policy actions as described in the following. $\mathbf{GB}_\mathrm{FCS}$ are the adaptive gain biases that modify the scheduled gains for the three-loop FCS. These biases are integrated from the policy actions, and must be included in the observation space. The observations are scaled to have zero mean and unit variance.

$$\mathbf{o} = \begin{bmatrix} \alpha_\mathrm{obs} & \beta_\mathrm{obs} & M_\mathrm{obs} & h_\mathrm{obs} & \boldsymbol{\omega}_\mathrm{obs} & \mathbf{a}^\mathrm{B}_\mathrm{err} & \boldsymbol{\delta}_\mathrm{C} & \mathbf{GB}_\mathrm{FCS} \end{bmatrix} \tag{20}$$

The output of the policy $\mathbf{u} = \pi(\mathbf{o}) \in \mathbb{R}^{12}$ is scaled through multiplication by $u_\mathrm{scale}$ and input to an integrator that saturates at $u_\mathrm{limit}$. The integrator output is then split into two vector components, each in $\mathbb{R}^6$, with each set biasing the pitch and yaw channel three loop autopilot gains. The integration improves performance by smoothing the action. We use $u_\mathrm{scale} = 18$ and $u_\mathrm{limit} = 0.3$

The reward function is shown below in Equations (21a) through (21e). $r_\mathrm{shaping}$ is a shaping reward given at each step in an episode. These shaping rewards take the form of a Gaussian-like function of the norm of the scaled acceleration tracking error as shown in Eq. (21a). $r_\mathrm{ctrl}$ is a control effort penalty, again given at each step in an episode, and $r_\mathrm{bonus}$ is a bonus given at the end of an episode if the miss distance is below the threshold $r_\mathrm{lim}$. Importantly, the current episode is terminated if a path constraint is violated, in which case the stream of positive shaping rewards is terminated, and the agent receives a negative reward. We use $\alpha = 2$, $\beta = -0.05$, $\delta = -0.10$, $\epsilon = 10$, $\zeta = -10$, $r_\mathrm{lim} = 3$ m, and $\sigma_\mathrm{acc} = 0.2$. We use a discount rate of 0.90 for shaping rewards and 0.995 for the terminal reward.

$$r_\mathrm{shaping} = \alpha \exp\left(\frac{\left(\|\mathbf{a}^\mathrm{B}_\mathrm{err}\|/a_\mathrm{max}\right)^2}{\sigma^2_\mathrm{acc}}\right) \tag{21a}$$

$$r_\mathrm{ctrl} = \delta \mathbf{u} \tag{21b}$$

$$r_\mathrm{bonus} = \begin{cases} \epsilon, & \text{if } \mathbf{r}_\mathrm{TM} < r_\mathrm{lim} \text{ and done} \\ 0, & \text{otherwise} \end{cases} \tag{21c}$$

$$r_\mathrm{penalty} = \begin{cases} \zeta, & \text{if any path constraint violated} \\ 0, & \text{otherwise} \end{cases} \tag{21d}$$

$$r = r_\mathrm{shaping} + r_\mathrm{ctrl} + r_\mathrm{bonus} + r_\mathrm{penalty} \tag{21e}$$

The gain definitions are from [9], and defined in Table 3. Figure 10 illustrates the gain adaptation during an engagement with a jinking target. Note that the occasional bias saturation indicates that further performance improvement may have been possible with further hyperparameter tuning.

**Table 3 Gain Definitions**

| Gain | Definition |
|---|---|
| $K_{DC}$ | Steady State |
| $K_A$ | Acceleration feedback |
| $\omega_I$ | Forward path integrator |
| $K_R$ | Rate feedback |
| $K_1$ | Aerodynamic acceleration |
| $K_3$ | Aerodynamic body rate |

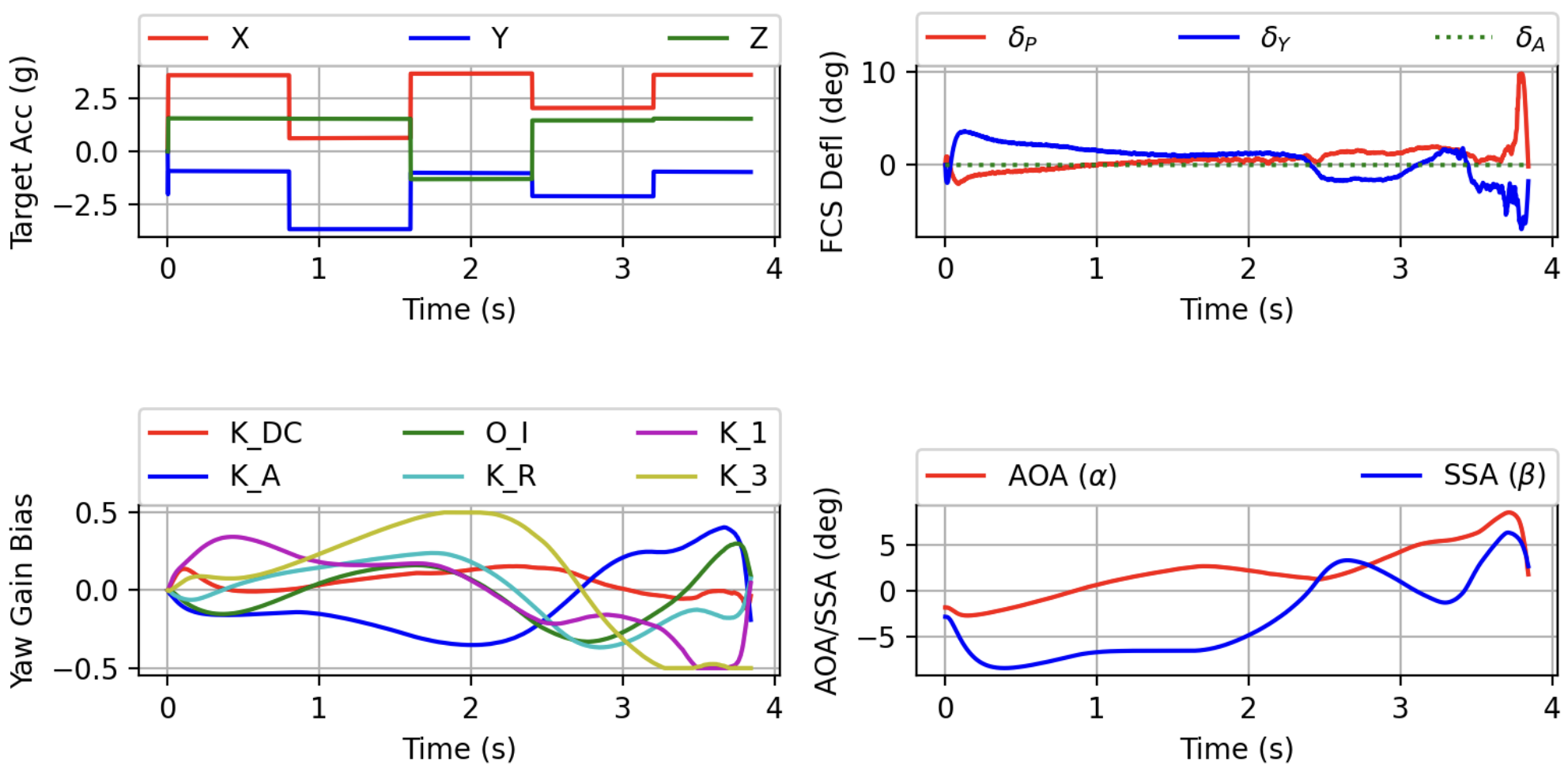

**Fig. 10 Adaptive Gain Biases over Engagement**

For optimization, aerodynamic uncertainty is modeled using a maximum coefficient bias of 10% and a maximum center of pressure shift equal to 1% (see Section II.D). A learning curve plotting the mean reward over rollouts versus episode is shown in Fig. 11. Here each optimization rollout consists of 30 simulated episodes. The reward history is slightly noisy due to the fact that the exploration variance does not fall to zero during the optimization (it would be destabilizing to the PPO algorithm if it did fall to zero). Note that for testing, the exploration variance is disabled, and we return the mean value of the policy distribution. The maneuver distribution used for both optimization and testing is 33% step, 33% weave, and 34% jinking.

### A. Meta-RL Problem Formulation, Full FCS (FCS-RL)

The agent observation $\mathbf{o}$ is shown in Eqs. (22), where $\alpha_{\text{obs}}$, $\beta_{\text{obs}}$, $M_{\text{obs}}$, $h_{\text{obs}}$, $\boldsymbol{\omega}_{\text{obs}}$, and $\mathbf{a}_{\text{err}}$, are generated as shown Sections II.C.3), II.C.4, and II.C.6. $\theta_{\text{roll}}$ is obtained through integration of the x component of the measured rotational velocity ($\omega_{\text{obs}}$) and $\boldsymbol{\delta}_{\text{C}}$ are the commanded channel deflections from Section II.C.8 and derived from policy actions as described in the following. The observations are scaled to have zero mean and unit variance.

$$\mathbf{o} = \begin{bmatrix} \alpha_{\text{obs}} & \beta_{\text{obs}} & M_{\text{obs}} & h_{\text{obs}} & \boldsymbol{\omega}_{\text{obs}} & \mathbf{a}_{\text{err}}^{\text{B}} & \boldsymbol{\delta}_{\text{C}} & \theta_{\text{roll}_{\text{obs}}} \end{bmatrix} \tag{22}$$

The output of the policy $\mathbf{u} = \pi(\mathbf{o}) \in \mathbb{R}^3$ is split into three scalar components, a pitch channel deflection rate $\dot{\delta}_{\text{P}}$, a yaw channel deflection rate $\dot{\delta}_{\text{Y}}$, and a roll channel deflection rate $\dot{\delta}_{\text{A}}$. These are scaled by multiplication with the maximum allowed deflection rate and then clipped to fall within the allowed commanded deflection rates. The maximum

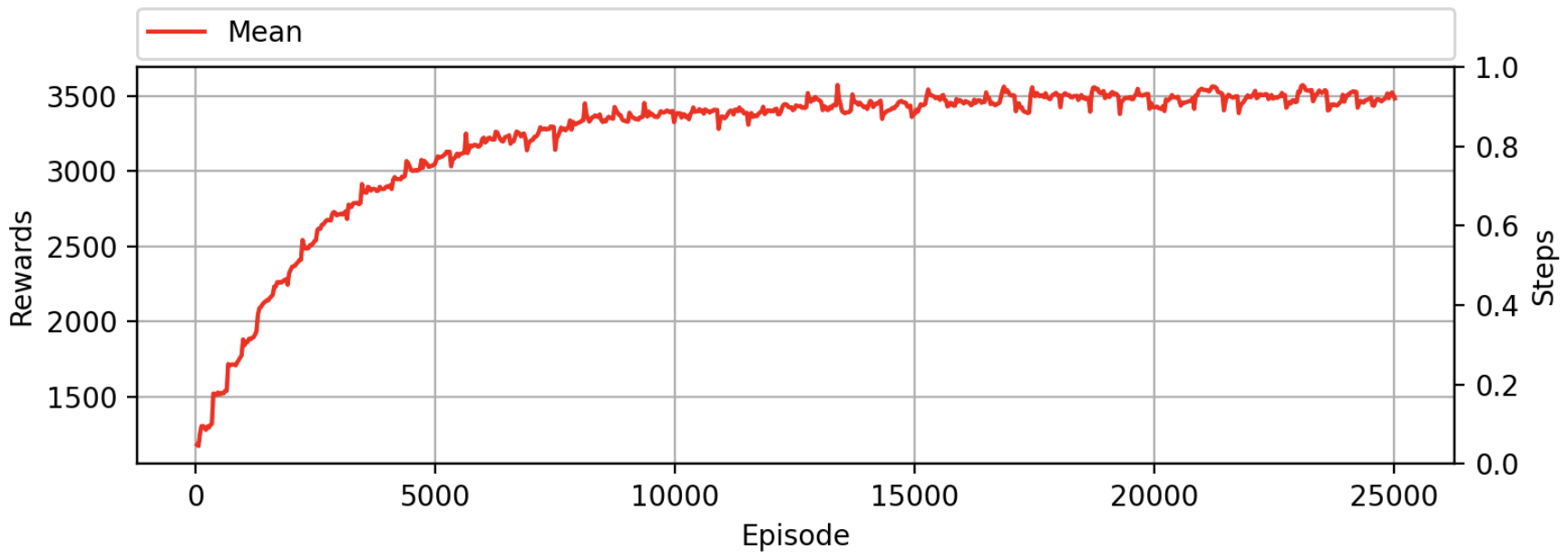

**Fig. 11 Optimization Reward History for FCS-3L-RL**

deflection rates are 500°/s. These commanded deflection rates are then integrated, with the integrator saturating at the maximum allowed deflection, giving the commanded deflections $\delta_{\text{P}}$, $\delta_{\text{Y}}$, and $\delta_{\text{A}}$

The reward function is shown below in Equations (23a) through (23f). $r_{\text{shaping}}$ is a shaping reward given at each step in an episode. These shaping rewards take the form of a Gaussian-like function of the norm of the scaled acceleration tracking error as shown in Eq. (21a). $r_{\text{roll}}$ encourages the agent to minimize deviations of body roll from zero, with differential fin deflections only used to counteract the rolling moment induced by angle of attack and sideslip (See Section II.D). $r_{\text{ctrl}}$ is a control effort penalty, again given at each step in an episode, and $r_{\text{bonus}}$ is a bonus given at the end of an episode if the miss distance is below the threshold $r_{\text{lim}}$. Importantly, the current episode is terminated if a path constraint is violated, in which case the stream of positive shaping rewards is terminated, and the agent receives a negative reward. We use $\alpha = 2$, $\beta = -0.05$, $\delta = -0.10$, $\epsilon = 10$, $\zeta = -10$, $r_{\text{lim}} = 3$ m, $\sigma_{\Omega} = 0.02$, and $\sigma_{\text{acc}} = 0.2$. We use a discount rate of 0.90 for shaping rewards and 0.995 for the terminal reward.

$$r_{\text{shaping}} = \alpha \exp\left(\frac{\left(\|\mathbf{a}_{\text{err}}^{\text{B}}\|/a_{\text{max}}\right)^2}{\sigma_{\text{acc}}^2}\right) \tag{23a}$$

$$r_{\text{roll}} = \beta|\theta_{\text{roll}}| \tag{23b}$$

$$r_{\text{ctrl}} = \delta \left\|\begin{bmatrix} \frac{\text{d}}{\text{d}t}(\delta_{\text{P}}) & \frac{\text{d}}{\text{d}t}(\delta_{\text{Y}}) & \frac{\text{d}}{\text{d}t}(\delta_{\text{A}}) \end{bmatrix}\right\| \tag{23c}$$

$$r_{\text{bonus}} = \begin{cases} \epsilon, & \text{if } \mathbf{r}_{\text{TM}} < r_{\text{lim}} \text{ and done} \\ 0, & \text{otherwise} \end{cases} \tag{23d}$$

$$r_{\text{penalty}} = \begin{cases} \zeta, & \text{if any path constraint violated} \\ 0, & \text{otherwise} \end{cases} \tag{23e}$$

$$r = r_{\text{shaping}} + r_{\text{ctrl}} + r_{\text{roll}} + r_{\text{bonus}} + r_{\text{penalty}} \tag{23f}$$

We use curriculum learning [35] to optimize a policy that adapts to high levels of aerodynamic uncertainty. Specifically, we first optimize the policy using coefficient bias bounds (CB) of 10% and center of pressure location bounds (CP) of 1% (see Section II.D). The resulting policy is then re-optimized using CB=20% and CP=2%, but with the policy initial exploration standard deviation set to 0.30 (as opposed to 1.00 for the initial optimization), allowing the second optimization run to retain learning from the first optimization, but giving the policy enough exploration ability to improve performance under the higher aerodynamic uncertainty. We then repeat this incremental optimization a third time, using CB=40% and CP=2%. This results in improved adaptability to aerodynamic uncertainty compared to trying a single full optimization using CB=40% and CP=2%. The reward learning curve for the initial optimization is shown in Fig. 12, where "Mean" is the mean reward over a set of rollouts. Before episode 10000, the agent is primarily learning to avoid constraint violations, which result in early termination of an episode. The maneuver distribution used for both optimization and testing is 33% step, 33% weave, and 34% jinking.

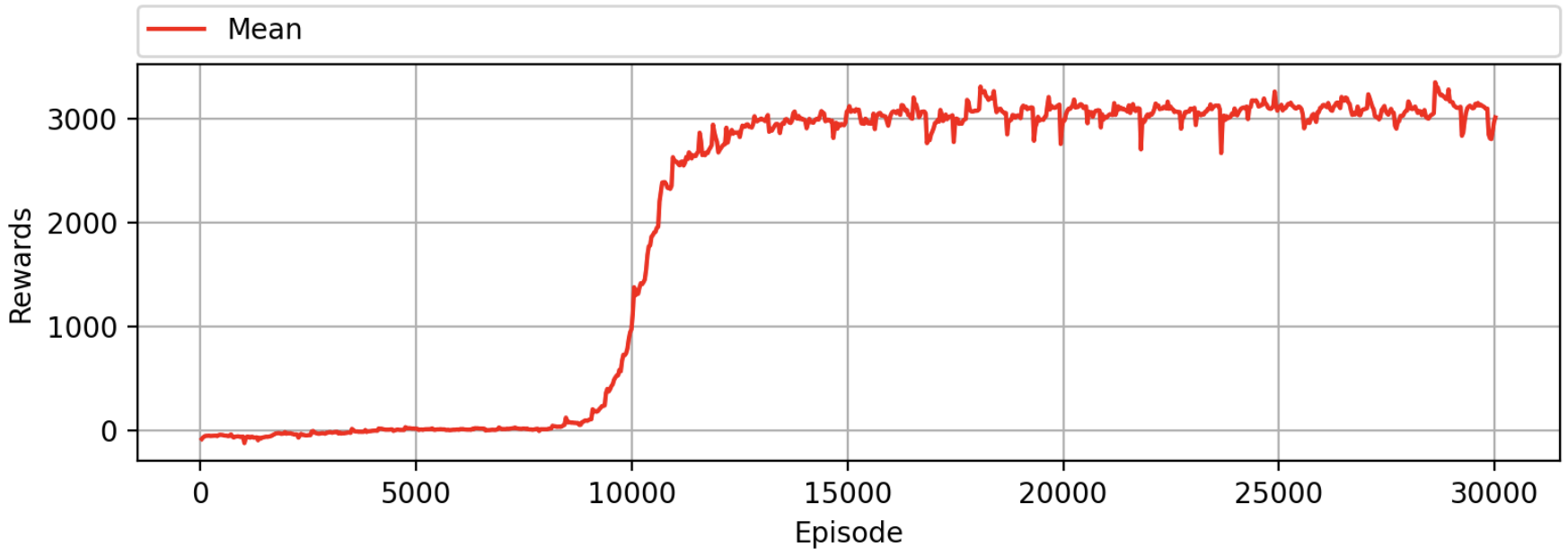

**Fig. 12 Optimization Reward History**

# VI. Experiments

## A. FCS Relative Performance

Relative performance is measured by running 5000 simulations for each case in the following tables. In the following tables, "AREF" (in g) is the reference acceleration (see Section II.B) for the target, "CB" is the aerodynamic coefficient bias, "CP" is the aerodynamic center of pressure shift (see Section II.D), and "R" is the maximum radome refraction in radians (see Section II.C.1. The FCS-RL, FCS-3L, and FCS-3L-RL use the labels 'RL', '3L', and '3L-A', respectively. Relative performance is quantified using several metrics. Accuracy is quantified by "Miss < 25 cm", which indicates the percentage of episodes that resulted in the interceptor achieving a miss distance less than or equal to 25 cm. "IR" is the interceptor ratio, which denotes the number of interceptors using the FCS-3L and FCS-3L-RL that are required to match the accuracy of an interceptor using the FCS-RL, according to the equation IR $= \dfrac{\log m_1}{\log m_2}$, where $m_1$ and $m_2$ are the probability of a miss distance greater than 25 cm for the FCS-RL and either FCS-3L or FCS-3L-RL, respectively. Control activity is measured using the root-mean-square of the magnitude of the interceptor acceleration "RMS $\|\mathbf{a}_\mathrm{M}\|$". Finally, "Vio (%)" gives the percentage of episodes that resulted in a load constraint violation. Table 4 shows that under conditions of no radome refraction, no aerodynamic uncertainty, and no target evasive acceleration, there is essentially no performance difference between the three flight control systems. To demonstrate the need for a recurrent policy in this application, we optimized a version of FCS-RL with the recurrent layer replaced with a standard feed forward layer (RL-MLP). The results clearly demonstrate the need for a recurrent policy (which allows realization of an integrator) when applying RL to flight control over a large flight envelope.

**Table 4 Ideal Conditions**

| Case | FCS | Miss < 25 cm (%) | IR | RMS $\|\mathbf{a}_\mathrm{M}\|$ (m/s$^2$) | Vio (%) |
|---|---|---|---|---|---|
| AREF=0 CB=0.0-CP=0.0 R=0 | **3L** | 100 | 1.0 | 41 | 0.0 |
| AREF=0 CB=0.0-CP=0.0 R=0 | **3L-RL** | 100 | 1.0 | 41 | 0.0 |
| AREF=0 CB=0.0-CP=0.0 R=0 | **RL** | 100 | - | 40 | 0.0 |
| AREF=0 CB=0.0-CP=0.0 R=0 | **RL-MLP** | 6 | - | 144 | 34.1 |

Table 5 tabulates performance when we consider radome refraction (Section II.C.1). Radome refraction results in parasitic interaction between the guidance and flight control systems that reduces performance. Except for the special case of a hemispherical radome, the actual LOS refraction depends on where the LOS passes through the radome. Furthermore, in order for the FCS to track a commanded acceleration, an interceptor using control surface actuation must rotate. This rotation results in the LOS to the target passing through a different part of the radome, causing the refraction angle to change, resulting in a false sense of target motion. The guidance system then reacts to this by giving a commanded acceleration that takes into account the false target motion, completing the parasitic feedback loop. We see that even with small radome refraction a performance differential appears between the flight control systems, with the differential growing as the radome refraction is increased.

For the simulations used to create Table 6 we increase AREF from 4 to 8, doubling the target's acceleration at a given dynamic pressure. Again, there is a clear difference between the performance of the three flight control systems. Next, Table 7 explores which target maneuvers are more difficult for the flight control systems. Whereas the performance for all flight control systems is nearly identical for the step maneuver, we see that a large performance differential exists for weaving and jinking maneuvers. Importantly, these types of maneuvers are likely to be employed by a target that is actively maneuvering to avoid interception, as it keeps the threat from deviating too far from the trajectory required to hit its target. Moreover, it is possible for certain types of reentry vehicles to spiral even without the capability of actively evading interception[36, 37].

**Table 5    Radome Sensitivity**

| Case | FCS | Miss < 25 cm (%) | IR | RMS $\|\|\mathbf{a}_M\|\|$ (m/s$^2$) | Vio (%) |
|---|---|---|---|---|---|
| AREF=4 CB=0.10-CP=0.01 R=0.0005 | **3L** | 70 | 2.5 | 61 | 0.0 |
| AREF=4 CB=0.10-CP=0.01 R=0.0005 | **3L-A** | 80 | 1.9 | 59 | 0.0 |
| AREF=4 CB=0.10-CP=0.01 R=0.0005 | **RL** | 95 | - | 56 | 0.0 |
| AREF=4 CB=0.10-CP=0.01 R=0.001 | **3L** | 60 | 2.5 | 60 | 0.0 |
| AREF=4 CB=0.10-CP=0.01 R=0.001 | **3L-A** | 70 | 1.9 | 59 | 0.0 |
| AREF=4 CB=0.10-CP=0.01 R=0.001 | **RL** | 90 | - | 56 | 0.0 |
| AREF=4 CB=0.10-CP=0.01 R=0.002 | **3L** | 35 | 2.7 | 63 | 0.0 |
| AREF=4 CB=0.10-CP=0.01 R=0.002 | **3L-A** | 45 | 2.0 | 62 | 0.0 |
| AREF=4 CB=0.10-CP=0.01 R=0.002 | **RL** | 69 | - | 57 | 0.0 |
| AREF=4 CB=0.10-CP=0.01 R=0.005 | **3L** | 11 | 2.7 | 66 | 0.0 |
| AREF=4 CB=0.10-CP=0.01 R=0.005 | **3L-A** | 15 | 1.9 | 65 | 0.0 |
| AREF=4 CB=0.10-CP=0.01 R=0.005 | **RL** | 27 | - | 63 | 0.0 |

**Table 6    Increased Target Evasive Acceleration (R=0.0005)**

| Case | FCS | Miss < 25c m (%) | IR | RMS $\|\|\mathbf{a}_M\|\|$ (m/s$^2$) | Vio (%) |
|---|---|---|---|---|---|
| AREF=8 CB=0.10-CP=0.01 | **3L** | 52 | 2.4 | 95 | 0.0 |
| AREF=8 CB=0.10-CP=0.01 | **3L-A** | 53 | 2.3 | 96 | 0.0 |
| AREF=8 CB=0.10-CP=0.01 | **RL** | 83 | - | 91 | 0.0 |

**Table 7    Target Maneuver Sensitivity (AREF=4, CB=0.10, CP=0.01, R=0.0005)**

| Case | FCS | Miss < 25 cm (%) | IR | RMS $\|\|\mathbf{a}_M\|\|$ (m/s$^2$) | Vio (%) |
|---|---|---|---|---|---|
| Step | **3L** | 98 | 1.2 | 65 | 0.0 |
| Step | **3L-A** | 97 | 1.3 | 65 | 0.0 |
| Step | **RL** | 99 | - | 46 | 0.0 |
| Weave | **3L** | 47 | 4.7 | 54 | 0.0 |
| Weave | **3L-A** | 65 | 2.9 | 52 | 0.0 |
| Weave | **RL** | 95 | - | 47 | 0.0 |
| Jink | **3L** | 66 | 2.7 | 63 | 0.0 |
| Jink | **3L-A** | 75 | 1.7 | 61 | 0.0 |
| Jink | **RL** | 90 | - | 57 | 0.0 |

Table 8 reports performance with various target weave periods, ranging from 0.1 to 1.0 seconds. We see that the weave period that maximizes miss distance is different between the three flight control systems. To see why, consider that in [38] an expression is derived showing peak miss distance as a function of normalized target weave frequency, specifically $\omega_T T$, where $\omega_T$ and $T$ are the target weave frequency and guidance system time constant, respectively. When taking into account FCS dynamics, the peak miss distance becomes a function of the effective time constant of the combined guidance and control system. Indeed, for a given guidance system, a decrease in flight control response time can improve effectiveness against weaving targets[38]. Importantly, although all three flight control system's performance are impacted by weaving targets, there is again a clear performance gap between the three systems. The performance difference is at least partly due to the improved flight control response time of the FCS-RL (See Fig 13 in the following). It is also possible that the FCS-RL is dynamically adapting its effective bandwidth to increase performance.

**Table 8 Target Weave Period (AREF=4, CB=0.10, CP=0.01, R=0.0005)**

| Weave Period (s) | FCS | Miss < 25 cm (%) | IR | RMS $\|\|\mathbf{a}_M\|\|$ (m/s$^2$) | Vio (%) |
|---|---|---|---|---|---|
| 0.1 | **3L** | 100 | 1.0 | 42 | 0.0 |
| 0.1 | **3L-A** | 100 | 1.0 | 41 | 0.0 |
| 0.1 | **RL** | 100 | - | 40 | 0.0 |
| 0.50 | **3L** | 50 | 1.4 | 45 | 0.0 |
| 0.50 | **3L-A** | 45 | 1.7 | 48 | 0.0 |
| 0.50 | **RL** | 63 | - | 45 | 0.0 |
| 1.00 | **3L** | 24 | 7.7 | 53 | 0.0 |
| 1.00 | **3L-A** | 30 | 5.9 | 53 | 0.0 |
| 1.00 | **RL** | 88 | - | 45 | 0.0 |

Table 9 illustrates the impact of increasing the level of aerodynamic uncertainty. We see that although all three FCS are robust to aerodynamic uncertainty, the performance advantage of both the FCS-RL and FCS-3L-RL are maintained as compared to FCS-3L, with a higher hit rate, lower control effort, and fewer constraint violations Finally, Table 10 gives interceptor and target statistics computed over 5000 trajectories, where we see that the acceleration of the interceptor and target are quite close at the end of the engagement, making this a highly agile target. Since the target starts at a much higher altitude than the interceptor, the lower dynamic pressure gives the target less maneuverability earlier in the engagement.

**Table 9 Aerodynamic Uncertainty (R=0.0005)**

| Case | FCS | Miss < 25 cm (%) | IR | RMS $\|\|\mathbf{a}_M\|\|$ (m/s$^2$) | Vio (%) |
|---|---|---|---|---|---|
| AREF=4 CB=0.10-CP=0.01 | **3L** | 70 | 2.5 | 61 | 0.0 |
| AREF=4 CB=0.10-CP=0.01 | **3L-A** | 80 | 1.9 | 59 | 0.0 |
| AREF=4 CB=0.10-CP=0.01 | **RL** | 95 | - | 56 | 0.0 |
| AREF=4 CB=0.20-CP=0.02 | **3L** | 66 | 2.5 | 62 | 0.6 |
| AREF=4 CB=0.20-CP=0.02 | **3L-A** | 77 | 1.8 | 60 | 0.4 |
| AREF=4 CB=0.20-CP=0.02 | **RL** | 93 | - | 57 | 0.0 |
| AREF=4 CB=0.40-CP=0.03 | **3L** | 60 | 1.9 | 65 | 3.4 |
| AREF=4 CB=0.40-CP=0.03 | **3L-A** | 70 | 1.5 | 63 | 3.2 |
| AREF=4 CB=0.40-CP=0.03 | **RL** | 84 | - | 60 | 0.8 |

Step response plots at 25 km altitude and 2500 m/s comparing the FCS-RL and FCS-3L under nominal, "slow", and "fast" aerodynamic models are shown in Fig. 13. The slow model has a reduction in aerodynamic force and moment coefficients and a shift in CP giving increased static margin, whereas the fast model has an increase in aerodynamic coefficients and reduced static margin. Although this is modeled as a shift in CP, it would be equivalent to a shift in CG in the opposite direction. We see that the FCS-RL exhibits a significant reduction in flight control response time as compared to the benchmark. A full trajectory for the FCS-RL starting at 25 km altitude with 0° inclination to target is shown in Fig. 14.

**Table 10 Typical High Altitude Engagement Statistics**

| | |
|---|---|
| Cumulative RMS $\lVert \mathbf{a}_{\mathrm{M}}^{\mathrm{B}}[y,z] \rVert$ (m/s$^2$) | 40 |
| Cumulative RMS $\lVert \mathbf{a}_{\mathrm{T}} \rVert \perp \boldsymbol{\lambda}_{\mathrm{TM}}$ (m/s$^2$) | 12 |
| Terminal RMS $\lVert \mathbf{a}_{\mathrm{M}}^{\mathrm{B}}[y,z] \rVert$ (m/s$^2$) | 58 |
| Terminal RMS $\lVert \mathbf{a}_{\mathrm{T}} \rVert \perp \boldsymbol{\lambda}_{\mathrm{TM}}$ (m/s$^2$) | 51 |
| Maximum Terminal Altitude $h$ (m) | 31516 |
| Minimum Terminal Altitude $h$ (m) | 12838 |

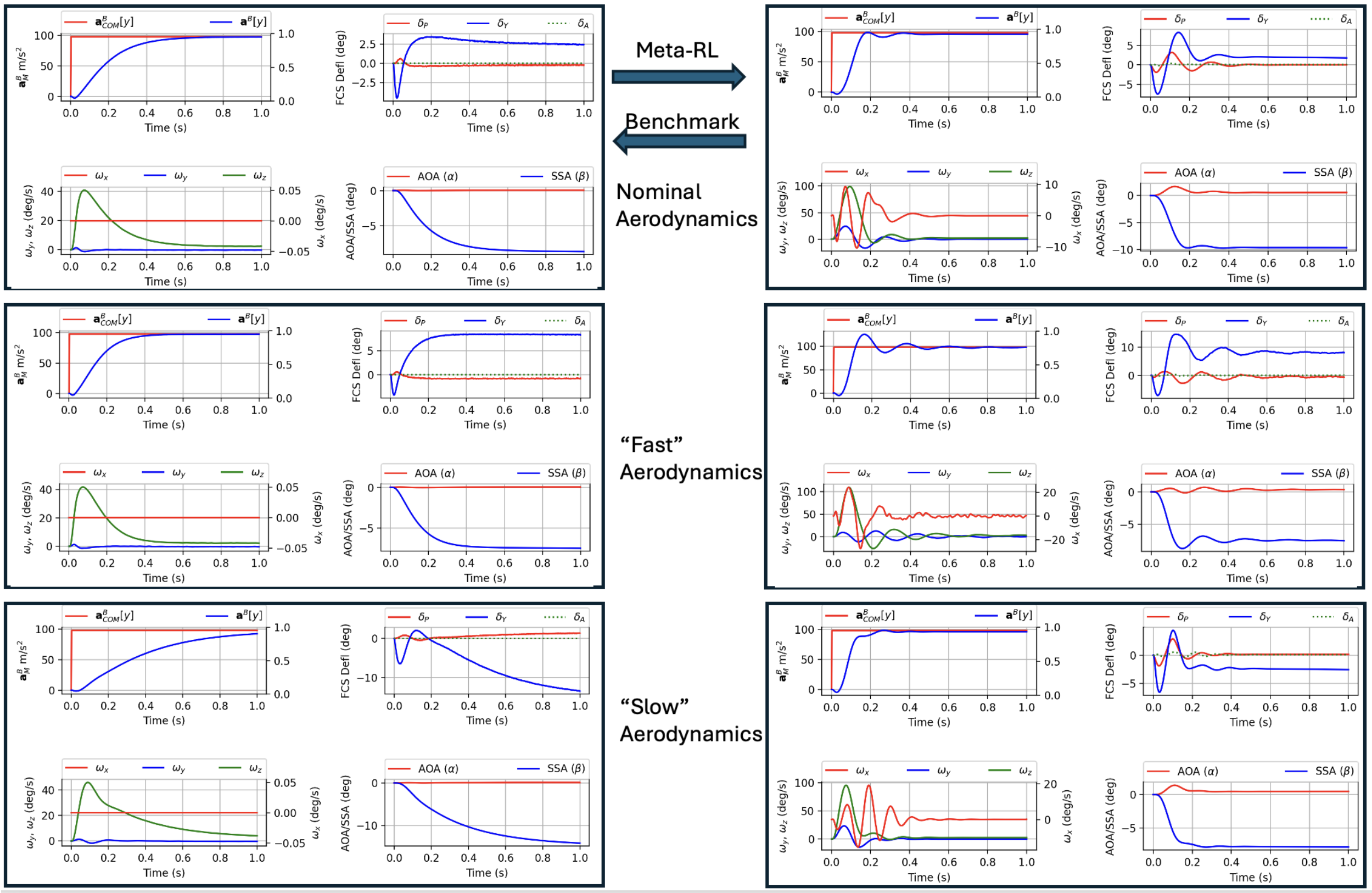

**Fig. 13 Step Response Comparison: 25 km, 2500 m/s**

## B. Policy Generalization Testing

We demonstrate the generalization capability of the three flight control systems by modifying the engagement initial conditions, as shown in Table 11. Note that in any real application, the homing phase is effective over some "basket" of initial conditions consistent with the end of the mid-course guidance phase. This basket is limited by a minimum range-to-target that gives the interceptor enough time to complete a successful terminal phase intercept, with the minimum time being a function of target maneuver, closing speed, and initial heading error. Using these modified initial conditions, we test the same target maneuver distribution as in Section VI.A, and also test against two target maneuver types not seen during optimization, a vertical S maneuver [39] (vert-S) and frequency modulated weave (FM-weave). The vertical S maneuver is identical to the weave maneuver except that the target acceleration instantaneously transitions between minimum and maximum acceleration at the weave maneuver zero acceleration crossings. In our simulator the vertical S differs from the jinking maneuver in that the jinking maneuver does not switch sign with a fixed period, and the jinking is independent for each axis. The frequency modulated weave has a random variation in the weave frequency over each engagement. This would be a relatively simple maneuver to execute for a maneuverable re-entry vehicle that is already spiraling with the fin deflections used to modulate the spiraling frequency.

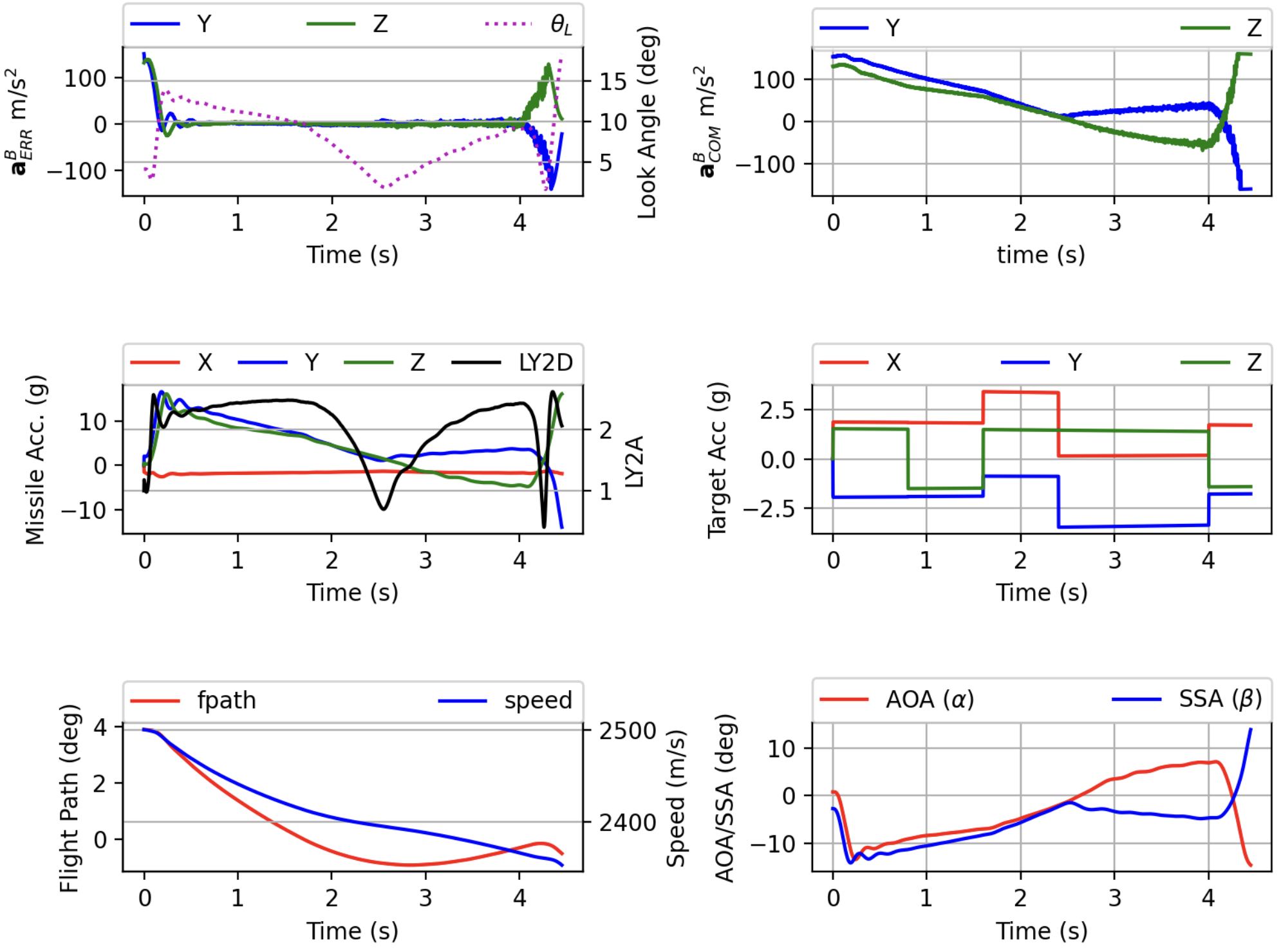

**Fig. 14 FCS-RL sample trajectory**

Example target acceleration trajectories for the FM-weave maneuver are illustrated in Fig. 15

**Table 11 Simulation Modified Initial Conditions)**

| | | |
|---|---|---|
| Elevation Angle to Target $\theta_E$ (degrees) | 30 | 60 |
| Azimuth Angle to Target $\theta_A$ (degrees) | -180 | 180 |
| Altitude (m) | 15000 | 22000 |
| Missile Velocity Magnitude $\|\mathbf{v}_\mathrm{M}\|$ (m/s) | 2200 | 2800 |
| Initial Range to Targeted Asset $\|\mathbf{r}_\mathrm{AT}\|$ (m) | 15000 | 30000 |
| Target Velocity Magnitude $\|\mathbf{v}_\mathrm{T}\|$ (m/s) | 2500 | 4500 |
| Target Velocity Cone Half Apex Angle $\theta_{\mathbf{v}_\mathrm{T}}$ (degrees) | 0 | 20 |

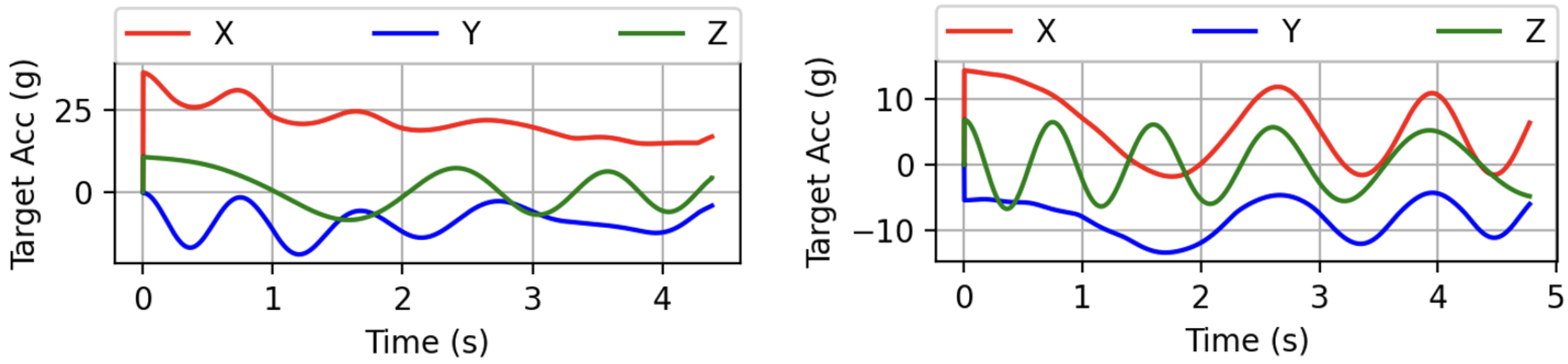

**Fig. 15 Frequency Modulated Weave Sample Trajectories**

From Table 12 we see that all three flight control systems are robust to an increase in both the range of initial conditions and uncertainty in the initial state estimate. The vertical-S maneuver is challenging for all three flight control systems, but more so for the FCS-3L and FCS-3L-RL. The key takeaway is that the RL FCS policy generalizes well to both initial conditions and target maneuvers outside of the optimization distribution. The FCS-RL performance is not affected by the FM-weave maneuver, whereas the FCS-3L and FCS-3L-RL are significantly reduced.

**Table 12 Generalization Cases**

| Case | UKF | Miss < 25cm (%) | RMS $\|\mathbf{a}_\mathrm{M}\|$ (m/s$^2$) | IR | Vio (%) |
|---|---|---|---|---|---|
| Mod-IC | **3L** | 73 | 49 | 1.9 | 0.0 |
| Mod-IC | **3L-A** | 83 | 49 | 1.4 | 0.0 |
| Mod-IC | **RL** | 92 | 46 | - | 0.0 |
| Mod-IC vert-S | **3L** | 42 | 49 | 2.2 | 0.0 |
| Mod-IC vert-S | **3L-A** | 47 | 58 | 1.9 | 0.0 |
| Mod-IC vert-S | **RL** | 70 | 45 | - | 0.0 |
| Mod-IC FM-weave | **3L** | 58 | 47 | 2.9 | 0.0 |
| Mod-IC FM-weave | **3L-A** | 54 | 52 | 3.3 | 0.0 |
| Mod-IC FM-weave | **RL** | 92 | 43 | - | 0.0 |

## VII. Flight Certification

Linear system theory can provide theoretical stability guarantees for the FCS-3L in the simplified case of a linearized system. However, optimal gain scheduling requires evaluating performance in a more realistic non-linear simulator [40]. Importantly, it is possible that for the case of a non-linear system, the system can be unstable at some flight condition while at the same time passing stability checks using the linearized system. This becomes more likely as the aerodynamics become more non-linear (which can be the case in the hypersonic flight regime). In addition, the actual deployment aerodynamics can be quite different from the simulated aerodynamics, weakening stability arguments based on sufficient gain or phase margin. Consequently, the FCS is not actually provably stable using linear systems theory when coupled with non-linear aerodynamics and significant aerodynamic uncertainty. Although a sufficiently large gain margin could compensate for differences between the actual and linearized systems, this comes with a performance cost. Consequently, extensive Monte Carlo simulation is required before moving to a flight test, which is also possible using the FCS-RL optimization process. Importantly, there has been recent work on the certification of stability for a meta-RL policy [41]. Finally, although the FCS-3L-RL underperformed the FCS-RL, it has the advantage of allowing flight certification using linear stability metrics, which could easily be implemented by limiting the gain bias to ensure that the resulting gains result in adequate margin for the linear system. This would allow a similar flight certification procedure as current systems.

## VIII. Conclusion

We applied reinforcement meta-learning (meta-RL) to two approaches to flight control system (FCS) optimization. In the first approach (FCS-RL) we optimize an adaptive policy directly mapping commanded acceleration tracking error and sensor outputs to commanded fin deflection rates in the pitch, yaw, and roll channels. In the second approach (FCS-3L-RL), the scheduled gains for a standard FCS (FCS-3L) using the three-loop autopilot for pitch and yaw control are adaptively adjusted during a trajectory using a meta-RL gain adaptation policy. The three flight control systems were then extensively tested in a 6-DOF simulator that modeled most factors relevant to flight control. Under ideal conditions, the performance of the three flight control systems was nearly identical. However, under conditions of radome refraction, target evasive maneuvers, and aerodynamic uncertainty, we found a significant performance differential appear in favor of the FCS-RL, with higher interception rates and lower control effort. We believe that there are two primary factors behind the FCS-RL performance. The first is the higher expressive power of the recurrent policy compared to a fixed three-loop topology, which allows for faster flight control response time, as measured by step response rise time. The second is the adaptability of the FCS-RL achieved by optimizing over an ensemble of aerodynamic models, and improved through curriculum learning. It is likely that the integrated pitch, yaw, and roll channel control has a smaller effect, but would have a much greater impact in bank-to-turn applications where aerodynamic coupling is higher. To put the performance differential in perspective, facing an actively evading target under small levels of radome refraction and aerodynamic uncertainty, the interception rate for the FCS-RL and FCS-3L are 95% and 70% respectively. This implies that 2.5 interceptors using the benchmark system would be required to reach the FCS-RL hit rate for a single interceptor. In other more challenging scenarios, the FCS-3L required more than 5 interceptors to match the FCS-RL performance. In addition to the obvious cost advantage, this increased accuracy can be critical in scenarios with a limited number of interceptors and/or launchers. Although FCS-3L-RL also demonstrates improved performance compared to FCS-3L-RL, the performance differential is smaller than FCS-RL. Finally, we demonstrate that both the FCS-3L-RL and the FCS-RL generalize well to initial conditions and target maneuvers that were not experienced during optimization.

# References

[1] Alviani, R., Poggie, J., and Blaisdell, G., "Unsteady Aspects of Shock-Wave/Boundary-Layer Interaction Resulting from Control Surface Deflection," *AIAA Journal*, Vol. 60, No. 8, 2022, pp. 4649–4659. https://doi.org/10.2514/1.J061337

[2] Edwards, J. R., "Numerical Simulations of Shock/Boundary Layer Interactions Using Time-Dependent Modeling Techniques: A Survey of Recent Results," *Progress in Aerospace Sciences*, Vol. 44, No. 6, 2008, pp. 447–465. https://doi.org/10.1016/j.paerosci.2008.06.004.

[3] Witeof, Z., and Pasiliao, C. L., "Fluid-Thermal-Structural Interaction Effects in Preliminary Design of High Speed Vehicles," *56th AIAA/ASCE/AHS/ASC Structures, Structural Dynamics, and Materials Conference*, 2015. https://doi.org/10.2514/6.2015-1631.

[4] Falkiewicz, N. J., Cesnik, C. E. S., Bolender, M. A., and Doman, D. B., "Thermoelastic Formulation of a Hypersonic Vehicle Control Surface for Control-Oriented Simulation," *AIAA Guidance, Navigation, and Control Conference*, 2009. https://doi.org/10.2514/6.2009-6284.

[5] Culler, A. J., and McNamara, J. J., "Impact of Fluid-Thermal-Structural Coupling on Response Prediction of Hypersonic Skin Panels," *AIAA Journal*, Vol. 49, No. 11, 2011, pp. 2393–2406. https://doi.org/10.2514/1.J050617.

[6] Zarchan, P., "Tactical and strategic missile guidance," American Institute of Aeronautics and Astronautics, Inc., 2012, pp. 126–132. https://doi.org/10.2514/4.868948.

[7] Siouris, G. M., "Missile guidance and control systems," Springer Science & Business Media, 2004, pp. 142–143. https://doi.org/10.1007/b97614.

[8] Shneydor, N. A., "Missile guidance and pursuit: kinematics, dynamics and control," Elsevier, 1998, pp. 151–155. https://doi.org/10.1533/9781782420590.

[9] Zarchan, P., "Tactical and strategic missile guidance," American Institute of Aeronautics and Astronautics, Inc., 2012, pp. 529–564. https://doi.org/10.2514/4.868948.

[10] Stallard, D. V., "An Approach to Autopilot Design for Homing Interceptor Missiles," *AIAA Guidance, Navigation, and Control Conference*, AIAA, 1991, pp. 99–113.

[11] Garnell, P., *Guided Weapon Control Systems*, 2nd ed., Pergamon Press, Oxford, 1980.

[12] Lee, H., Snyder, S., and Hovakimyan, N., "$L_1$ Adaptive Output Feedback Augmentation for Missile Systems," *IEEE Transactions on Aerospace and Electronic Systems*, Vol. 54, No. 2, 2018, pp. 680–692. https://doi.org/10.1109/TAES.2017.2764218.

[13] Rusnak, I., Weiss, H., and Barkana, I., "Improving the Performance of Existing Missile Autopilot Using Simple Adaptive Control," *International Journal of Adaptive Control and Signal Processing*, Vol. 28, No. 7-8, 2014, pp. 732–749. https://doi.org/10.1002/acs.2457.

[14] Joshi, S. M., Tao, G., and Khong, T., "A Direct Adaptive Control Approach in the Presence of Model Mismatch," *AIAA Guidance, Navigation, and Control Conference*, AIAA, Chicago, Illinois, 2009. https://doi.org/10.2514/6.2009-5620.

[15] Kannan, S. K., and Johnson, E. N., "Model Reference Adaptive Control with a Constrained Linear Reference Model," *49th IEEE Conference on Decision and Control*, IEEE, 2010, pp. 48–53. https://doi.org/10.1109/CDC.2010.5717275.

[16] Rohrs, C. E., Valavani, L., Athans, M., and Stein, G., "Robustness of Continuous-Time Adaptive Control Algorithms in the Presence of Unmodeled Dynamics," *IEEE Transactions on Automatic Control*, Vol. 30, No. 9, 1985, pp. 881–889. https://doi.org/10.1109/TAC.1985.1104070.

[17] Cao, C., and Hovakimyan, N., "Design and Analysis of a Novel $L_1$ Adaptive Control Architecture with Guaranteed Transient Performance," *IEEE Transactions on Automatic Control*, Vol. 53, No. 2, 2008, pp. 586–591. https://doi.org/10.1109/TAC.2007.914282.

[18] Shin, H.-S., He, S., and Tsourdos, A., "A Domain-Knowledge-Aided Deep Reinforcement Learning Approach for Flight Control Design," , 2019. URL https://arxiv.org/abs/1908.06884.

[19] Taligatos, R., Nguyen, D. H., and Lowenberg, M. H., "Investigating Missile Feedback Control: Reinforcement Learning, PID, or Both?" *AIAA SCITECH 2026 Forum*, AIAA, 2026. https://doi.org/10.2514/6.2026-1162.

[20] Candeli, A., De Tommasi, G., Lui, D. G., Mele, A., Santini, S., and Tartaglione, G., "A Deep Deterministic Policy Gradient Learning Approach to Missile Autopilot Design," *IEEE Access*, Vol. 10, 2022, pp. 19685–19696. https://doi.org/10.1109/ACCESS.2022.3150926.

[21] Wang, W., Wu, M., Chen, Z., and Liu, X., "Integrated Guidance-and-Control Design for Three-Dimensional Interception Based on Deep-Reinforcement Learning," *Aerospace*, Vol. 10, No. 2, 2023, p. 167. https://doi.org/10.3390/aerospace10020167.

[22] Gaudet, B., and Furfaro, R., "A Comparison of Partially and Fully Integrated Guidance and Flight Control Optimized with Reinforcement Meta-Learning," *AIAA SCITECH 2023 Forum*, 2023, p. 1628.

[23] Gaudet, B., Charcos, I., and Furfaro, R., "Integrated and Adaptive Guidance and Control for Endoatmospheric Missiles via Reinforcement Learning," *arXiv preprint arXiv:2109.03880*, 2021.

[24] Gaudet, B., Durant, D., and Furfaro, R., "Reinforcement Learning for Target Jerk Estimation in Interceptor Homing Phase," , 2026. URL https://arxiv.org/abs/2611.15923.

[25] Gaudet, B., Furfaro, R., Linares, R., and Scorsoglio, A., "Reinforcement Metalearning for Interception of Maneuvering Exoatmospheric Targets with Parasitic Attitude Loop," *Journal of Spacecraft and Rockets*, Vol. 58, No. 2, 2021, pp. 386–399.

[26] Shneydor, N. A., "Missile guidance and pursuit: kinematics, dynamics and control," Elsevier, 1998, p. 118. https://doi.org/10.1533/9781782420590.

[27] Vasile, J. D., Bryson, J. T., Sahu, J., Paul, J. L., and Gruenwald, B. C., "Aerodynamic Dataset Generation of a Long-Range Projectile," Tech. rep., ARMY RESEARCH LAB ABERDEEN PROVING GROUND MD ABERDEEN PROVING GROUND United . . . , 2020.

[28] Jorgensen, L. H., "A method for estimating static aerodynamic characteristics for slender bodies of circular and noncircular cross section alone and with lifting surfaces at angles of attack from 0 deg to 90 deg," 1973.

[29] Zarchan, P., "Tactical and strategic missile guidance," American Institute of Aeronautics and Astronautics, Inc., 2012, pp. 473–482. https://doi.org/10.2514/4.868948.

[30] Zarchan, P., "Tactical and strategic missile guidance," American Institute of Aeronautics and Astronautics, Inc., 2012, pp. 18–21. https://doi.org/10.2514/4.868948.

[31] Zarchan, P., "Advanced Tactical and Strategic Missile Guidance, Volume 2," American Institute of Aeronautics and Astronautics, Inc., 2012, pp. 1–29. https://doi.org/10.2514/4.105388.

[32] Furfaro, R., and Gaudet, B., "Deep Learning Framework for Rapid development of High Performance Hypersonisc Weapons, 2021-002 T2, GN&C Working Group Final Report," *University Consortium for Applied Hypersonics, Contract: HQ0034229CA10*, 2026. Final report is CUI, contact UCAH for access. Report will appear in DTIC. PI: Furfaro, Working group lead: Gaudet.

[33] Wang, J. X., Kurth-Nelson, Z., Tirumala, D., Soyer, H., Leibo, J. Z., Munos, R., Blundell, C., Kumaran, D., and Botvinick, M., "Learning to reinforcement learn," *arXiv preprint arXiv:1611.05763*, 2016.

[34] Schulman, J., Wolski, F., Dhariwal, P., Radford, A., and Klimov, O., "Proximal policy optimization algorithms," *arXiv preprint arXiv:1707.06347*, 2017.

[35] Narvekar, S., Peng, B., Leonetti, M., Sinapov, J., Taylor, M. E., and Stone, P., "Curriculum Learning for Reinforcement Learning Domains: A Framework and Survey," *Journal of Machine Learning Research*, Vol. 21, No. 181, 2020, pp. 1–50.

[36] Chadwick, W. R., and Zarchan, P., "Interception of Spiraling Ballistic Missiles," *Proceedings of the American Control Conference*, IEEE, Seattle, Washington, USA, 1995, pp. —. Discusses guidance and interception performance against spiraling ballistic reentry vehicles.

[37] Platus, D. H., "Ballistic Re-entry Vehicle Flight Dynamics," *Journal of Guidance and Control*, Vol. 5, No. 1, 1982, pp. 4–16. https://doi.org/10.2514/3.56131, analysis of ballistic reentry vehicle aerodynamic and rotational dynamics during atmospheric reentry.

[38] Zarchan, P., "Tactical and strategic missile guidance," American Institute of Aeronautics and Astronautics, Inc., 2012, pp. 439–471. https://doi.org/10.2514/4.868948.

[39] Zarchan, P., "Tactical and strategic missile guidance," American Institute of Aeronautics and Astronautics, Inc., 2012, pp. 120–121. https://doi.org/10.2514/4.868948.

[40] Jackson, P. B., "Overview of Missile Flight Control Systems," *Johns Hopkins APL Technical Digest*, Vol. 29, No. 1, 2010, pp. 9–24. URL https://www.jhuapl.edu/sites/default/files/2024-09/29-01-Jackson.pdf.

[41] Scorsoglio, A., D'Ambrosio, A., and Furfaro, R., "[Article Title]," *Journal of Guidance, Control, and Dynamics*, 2026, pp. 1–17. https://doi.org/10.2514/1.G009769. URL https://doi.org/10.2514/1.G009769, articles in Advance.